\pdfoutput=1

\documentclass[10pt,twocolumn,reprint,amsmath,amssymb,amsfonts,aps,prd,superscriptaddress,showpacs,nofootinbib]{revtex4-1}
\usepackage{graphicx}
\usepackage{hyperref}
\usepackage[normalem]{ulem}
\usepackage{slashed}
\usepackage{float}
\usepackage[caption=false]{subfig}
\usepackage{bm}
\usepackage[utf8]{inputenc}

\hypersetup{
	bookmarks=true,         
	unicode=true,          
	pdftoolbar=true,        
	pdfmenubar=true,        
	pdffitwindow=false,     
	pdfstartview={FitH},    
	pdftitle={SecondaryDP},    
	pdfauthor={Author},     
	pdfsubject={Subject},   
	pdfcreator={Creator},   
	pdfproducer={Producer}, 
	pdfkeywords={keyword1} {key2} {key3}, 
	pdfnewwindow=true,      
	colorlinks=true,       
	linkcolor=black,          
	citecolor=black,        
	filecolor=black,      
	urlcolor=black           
}

\newcommand{\be}{\begin{equation}}
\newcommand{\ee}{\end{equation}}
\def\bsp#1\esp{\begin{split}#1\end{split}}

\newcommand{\amc}{\texttt{MadGraph5\_aMC}$@$\texttt{NLO} }

\newcommand{\geant}{\texttt{\textsc{Geant4}} }
\newcommand{\geantnospace}{\texttt{\textsc{Geant4}}}

\def\nn{\nonumber}
\def\nez{n_e^0}
\def\ng{n_\gamma^0}
\def\ne{n_e}
\def\Ez{E_{\gamma}}
\def\aem{\alpha_{\rm em}}

\begin{document}
	
\title{ New production channels for light dark matter in hadronic showers}

\newcommand*{\INFNGE}{Istituto Nazionale di Fisica Nucleare, Sezione di Genova, 16146 Genova, Italy}	\newcommand*{\INFNFR}{Istituto Nazionale di Fisica Nucleare, Laboratori Nazionali di Frascati, C.P. 13, 00044 Frascati, Italy}

\author{A.~Celentano}
\affiliation{\INFNGE}
\author{L.~Darm\'e}
\affiliation{\INFNFR}
\author {L.~Marsicano} 
\affiliation{\INFNGE}
\author{E.~Nardi}
\affiliation{\INFNFR}

	\begin{abstract}
	Hadronic showers transfer a relevant amount of their energy to electromagnetic subshowers. 
	We show that the generation of ``secondary'' dark photons in these sub-showers is significant and typically dominates the production at low dark photon masses. The resulting dark photons are however substantially less energetic than the ones originating from mesons decay.
	We illustrate this point both semi-analytically and through Monte Carlo simulations. 
	Existing limits on vector-mediator scenarios for light dark matter are updated 
	with the inclusion of the new production processes.
	\end{abstract}
		\vspace{1cm}
		\maketitle
	\tableofcontents
	\setcounter{footnote}{0}

\section{Introduction}

One of the most compelling empirical arguments to search for  extensions 
of the Standard Model (SM) of elementary particles is the need to explain 
the nature of dark matter (DM). 
In years past,  theoretical and experimental efforts mainly catalysed  around the hypothesis 
that DM corresponds to  a Weakly Interacting Massive Particle (WIMP) with  
electroweak scale mass (for a recent review, see e.g. Ref.~\cite{Roszkowski:2017nbc}). 
Such a hypothesis is certainly well grounded, given that in the early Universe WIMPs would 
be produced via thermal processes, and their subsequent annihilation with typical weak interaction rates 
would  leave, almost independently of other details, a relic density of the correct size to 
match the observed cosmological amount of DM. However, null results of an extensive and 
long lasting search program that combined direct, indirect, and collider probes are 
presently triggering a waning of the WIMP paradigm~\cite{Arcadi:2017kky}.
While WIMP searches should certainly continue until all experimentally accessible corners of the  
parameter space are thoroughly probed, it is now timely and important to put no lesser vigor 
in exploring also other pathways.  

One alternative scenario, well motivated in first place by the evidence that DM is reluctant to interact with ordinary matter,  conjectures the existence of a new class of relatively light elementary particles not charged under the SM  $SU(3)_C\times SU(2)_L\times U(1)_Y$  gauge group. After all, even in the SM some particles are uncharged under one or more gauge group factors, so that an extension to include a new sector blind to all SM interactions is not particularly exotic. In addition,  given that in the SM there is no shortage of states with mass below, say,  1~GeV/c$^2$, it is also rather natural to hypothesise that the same could be true for dark sector particles, 
the lightest of which would be stable, thus providing a light dark matter (LDM) candidate.
On the other hand, the dark sector could well come equipped with its own set of interactions
(to which SM particles should clearly be blind) and if this set also contains the simplest 
type of gauge force,  corresponding to a $U(1)$ gauge factor, then mixing between the 
dark spin-1 boson (often referred to as ``dark photon'' and denoted as $V$ in this work)
and the photon would naturally occur~\cite{HOLDOM1986196}. 
This would provide a portal though which the SM and the dark sector could communicate. 

Recent years have witnessed a steadily growing interest towards LDM and its possible detection through the vector portal, and many  studies have appeared deepening our understanding of the theoretical models and of their phenomenology,  see for example Refs.~\cite{PhysRevD.88.114015,Batell:2009di,Izaguirre:2015yja,Knapen:2017xzo,Feng:2017drg,Fabbrichesi:2020wbt,Filippi:2020kii}. 
Interestingly, besides promoting  new experimental programs aiming to search both for the $V$ and for LDM particles~\cite{Battaglieri:2017aum,Essig:2013lka,Alexander:2016aln}, the LDM paradigm also stimulated the reanalysis and reinterpretation of old data originally collected 
to search for other types of particles~\cite{Batell:2009di,deNiverville:2011it,Batell:2014mga,Lees:2017lec}. Accelerator-based thick-target experiments at moderate beam energy ($\sim$ 10$\div$100 GeV) are the ideal tool to probe the new hypothesis, since they have a very large discovery potential in a wide area of parameters space. Within this context, the main experimental techniques that have been considered so far are (1) missing energy/momentum/mass searches with electron and/or positron beams~\cite{Lees:2017lec,NA64:2019imj,Akesson:2018vlm,Duerr:2019dmv}, (2) electron and proton thick-target experiments searching for light new particles via their scattering in a downstream detector~\cite{PhysRevD.88.114015,deNiverville:2016rqh}, and (3) decay of long-lived dark sector fields into SM particles~\cite{TuckerSmith:2001hy,Izaguirre:2017bqb,Darme:2017glc,Berlin:2018jbm,Berlin:2018pwi,Darme:2018jmx,Mohlabeng:2019vrz,Tsai:2019mtm,Jodlowski:2019ycu}.

Proton beam-dump experiments show an enhanced sensitivity to the dark sector. Thanks to the large beam energy and accumulated charge, typically higher than those in electrons and positrons counterparts, a large LDM signal yield is expected, usually at the price of a larger background~\cite{Batell:2014mga,Berlin:2018bsc,Battaglieri:2019nok}. The experimental intensity frontier is currently extremely active, and many new experiments will start to take data during the course of the next decade \cite{Beacham:2019nyx,Bauer:2018onh,Battaglieri:2017aum}, making the accurate estimation of their potential reaches an important issue. This is particularly true for dark sector searches carried out at proton beam-dump experiments designed for neutrino physics~\cite{deNiverville:2016rqh,deNiverville:2018dbu}, such as MiniBooNE \cite{AguilarArevalo:2008qa}, SBND \cite{Antonello:2015lea}, ICARUS \cite{Antonello:2013ypa} or DUNE~\cite{Abi:2018dnh}, and for lower energy COHERENT~\cite{Akimov:2018ghi,Dutta:2019nbn}, where the irreducible neutrino background calls for an even more careful evaluation of the expected LDM signal. 

In the aforementioned  vector portal scenario in which a light dark photon interacts with the SM sector 
via  feeble gauge interactions, the main LDM production mechanism involved in a proton beam-dump experiment is the two-photons decay of light mesons ($\pi^0$ and $\eta$), where dark sector particles are produced thanks to the $\gamma-V$ mixing.
This production mechanism has been widely studied in the last decades. However, a proton-induced hadronic shower is always accompanied by an electromagnetic counterpart, which carries a significant fraction of the primary beam energy. This allows for a rich variety of electron- and positron-induced LDM production processes, incrementing the flux of LDM particles from the thick target, and thus the experimental sensitivities. An early attempt to consider this effect was presented in~\cite{Gorbunov:2014wqa}, considering only the $V$ visible decay.
In this work, for the first time we estimate the LDM production rate from the electromagnetic components of proton beam-dump experiments. We show that, in some cases, this is the dominant LDM production mechanism for a non-negligible 
region of the dark sector parameter space. Furthermore,  we demonstrate that 
thanks to these new production processes 
proton beam-dump experiments can also probe non-minimal dark sector scenarios that were, so far, considered to be an unique prerogative of lepton-beam efforts, such as protophobic models~\cite{Battaglieri:2017aum} in which the dark photon coupling to quarks is strongly suppressed. The recently proposed protophobic fifth-force interpretation~\cite{Feng:2016jff,Feng:2020mbt}  of the observed anomalies in internal $e^\pm$ pair creation in $^{8}$Be and $^4$He nuclear transitions~\cite{Krasznahorkay:2015iga,Krasznahorkay:2019lyl} is an example of a 
particularly intriguing new physics scenario that the new production mechanisms allow to test also 
in proton beam thick-target experiments.

The paper is organised as follows. In Sec.~\ref{sec:LDMprod} we describe the phenomenology of LDM production by secondary electrons and positrons in a $\sim$ 100 GeV proton beam-dump experiment, discussing both the main properties of proton-induced  electromagnetic  showers and the dominant LDM processes induced by $e^+$ and $e^-$ at this energy scale. In Sec.~\ref{sec:numerics} we present the details of the numerical procedure that we developed to compute the enhanced sensitivity of proton beam-dump experiments, which  results from taking into account
the new production processes. Finally, in Sec.~\ref{sec:expresults}, after briefly reviewing the main features 
of a representative set of proton beam-dump experiments, we present the corresponding exclusion limits and 
sensitivity curves, updated by including the new LDM production channels.

\section{LDM production by secondary $e^\pm$ in proton beam-dump experiments}
\label{sec:LDMprod}
The production of dark-sector particles in a proton beam-dump experiment is a multi-step process involving the secondary particles produced in the thick target by the impinging hadron.
Due to the  variety of secondary particles being part of the developing hadronic shower, a large number of production mechanisms is possible. In this work, we include for the first time electron- and positron-induced processes in the computation of the LDM yield of proton beam-dump experiments. 
In order to do so, we decouple the problem into two separate parts: the development of the EM shower
which is controlled by SM physics, and the new physics processes generating the dark photon 
(which we assume to be strongly sub-dominant compared to the former). 
More in detail, we will first revisit the typical structure of the EM component of a proton-induced hadronic shower, for a primary beam energy in the 10$\div$100 GeV range. We will next discuss the main processes responsible for LDM production by electrons and positrons in this regime. Finally,  we will focus on the production and detection of LDM in a typical proton-beam, thick-target experiment.

\subsection{Production of $e^\pm$ in proton-induced hadronic showers} 

When a high-energy proton impinges on a thick target, a cascade of secondary hadrons with progressively degrading energy is produced,  mostly containing protons, neutrons, and pions. Due to the isospin symmetry of hadron-induced reactions, approximately 1/3 of the latter are $\pi^0$. These immediately decay to  high-energy $\gamma\gamma$ pairs, which in turn initiate an EM shower accompanying the hadronic one. A similar argument applies for  $\eta$ and $\eta^\prime$ mesons, although their contribution to the EM shower is reduced, both because of the smaller production cross section, and because of the lower branching fraction for the $\gamma \gamma$ decay. On average, the fraction of the primary proton energy transferred to the EM component is of the order 50$\%$ for a 100 GeV impinging proton~\cite{Fabjan:2003aq}.

\begin{figure}[tpb]
    \centering
    \includegraphics[width=0.48\textwidth]{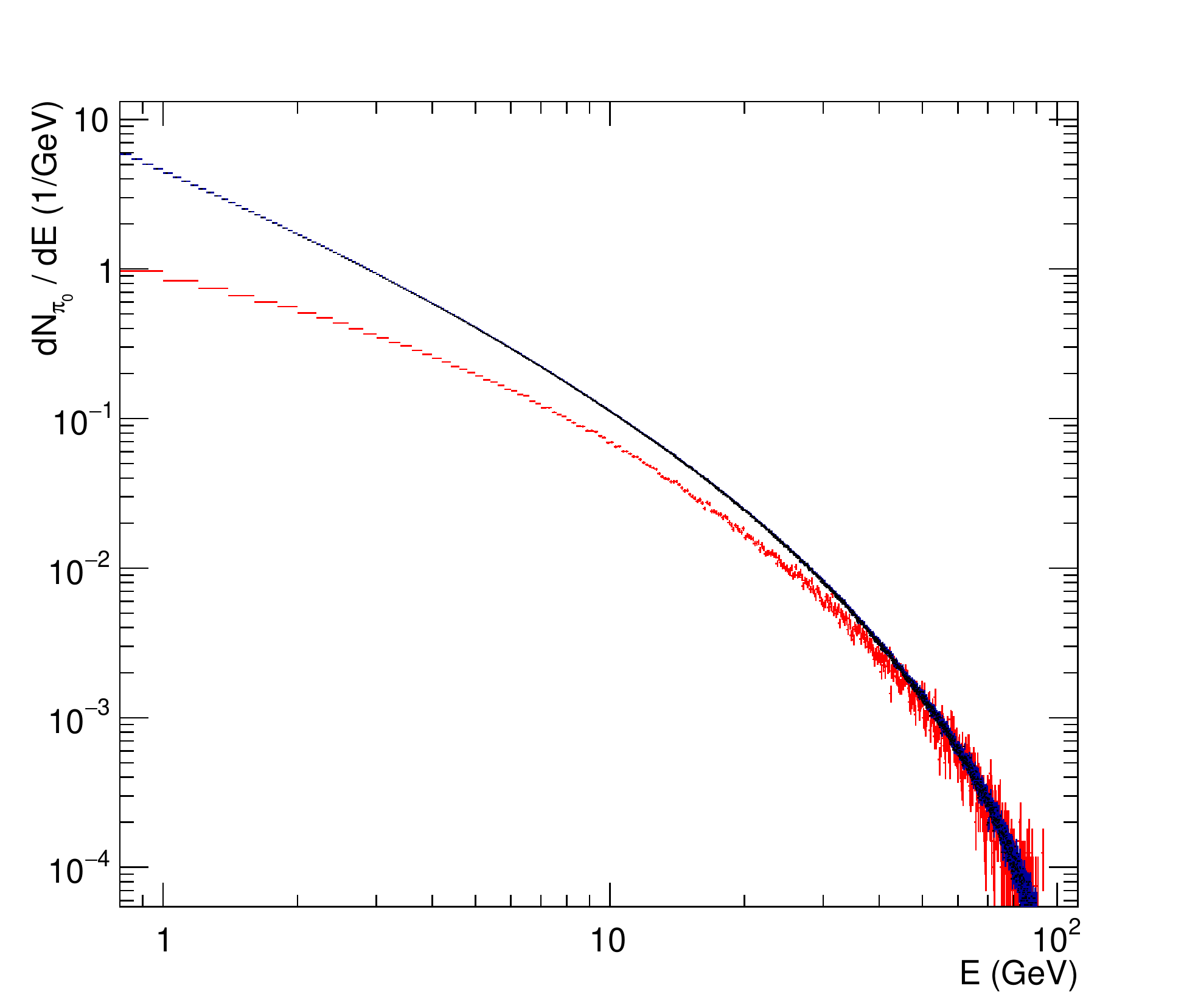}
    \caption{
    Comparison of the differential $\pi^0$ yield per proton on target from a 120 GeV proton impinging on a thick graphite target.
     Red curve: result obtained from Eq.~\ref{eq:neutmesonyield} based on \texttt{EPOS-LHC}~\cite{Pierog:2013ria}. Blue curve: results of a full \texttt{\geant}-based simulation.}
    \label{fig:ComparisonPi0}
\end{figure}

While a complete treatment  based on  numerical simulations will be presented in the next sections, a relatively good approximation of the energy distributions can be obtained from a semi-analytical approach. Starting from the typical differential number density of secondary neutral mesons $n_{M_0}(E)$ from a $pN$ collision 
 at the beam energy, 
with $N$ a nucleus of the target material, the differential yield of mesons in the hadronic shower, per POT, can be estimated approximately by just considering the first interaction of the proton:
\begin{align}\label{eq:neutmesonyield}
\frac{d N_{M_0}}{dE}  = \frac{\mathcal{N}_A \rho}{A} L \,\sigma_{pN} \times \frac{d n_{M_0}(E)}{dE} ~\equiv \frac{L}{\lambda_T} \times \frac{d n_{M_0}(E)}{dE}\ ,
\end{align} 
where $\sigma_{pN}$ is the inelastic proton-nucleon cross section, $A$ is the atomic mass of the target material, $\rho$ the density,  $\mathcal{N}_A  = 6.022 \times 10^{23}$, $L$ is the  length of the  active part of the target, and the  second equality follows from the definition of the nuclear interaction length $\lambda_T$. If the target is thick enough, $L \gtrsim \lambda_T$, and in the approximation of only considering the first generation of secondary particles in the hadronic shower, we can set  $L\simeq \lambda_T$, which results in the simplified expression   $\frac{dN_{M_0}}{dE} \simeq \frac{d n_{M_0}}{d E}$.
Clearly, this approximation is expected to be more accurate for energies close to the beam energy, while for lower energies the actual number of neutral mesons  would be underestimated. This effect is clearly visible in Fig.~\ref{fig:ComparisonPi0}, where we show the differential $\pi^0$ yield from a 120 GeV proton beam impinging on a thick graphite target, comparing the approximate result from Eq.~\ref{eq:neutmesonyield} (red curve) with that obtained from a full simulation of the hadronic shower made with \texttt{\geant}~\cite{AGOSTINELLI2003250} (blue curve). We used the \texttt{QGSPJETII} software~\cite{Ostapchenko:2010vb} to compute $\frac{d n_{M_0}}{d E}$ - a similar calculation with the \texttt{EPOS-LHC} software~\cite{Pierog:2013ria} yielded the same conclusion. From the knowledge of $\frac{dN_{M_0}}{dE}$, the differential distribution of primary photons from neutral mesons decay $\frac{dN_\gamma}{dE_\gamma}$ can be obtained, accounting for the corresponding branching fraction.

While a thorough description of the EM shower development requires a complete Monte Carlo calculation, an approximate evaluation of the electrons and positrons track length can still be obtained with an analytical approach. We introduce the dimensionless shower depth parameter in unit of radiation length $t \equiv d / X_0$ and the shower age as function of the energy $E$ and $t$~\cite{RevModPhys.13.240,Lipari:2008td}:
\begin{align}
\label{eq:showerage}
s\left(\frac{E}{\Ez},t \right) \simeq \frac{3t }{t - 2 \ln \frac{E}{\Ez} } \ ,
\end{align}
where $\Ez$ is the energy of the photon inducing the shower. The differential distribution $n_e$ of electron/positron grows exponentially with $s$, corresponding to the power law scaling
\begin{align}
    n_e \sim \frac{1}{E^{s+1}} \ .
\end{align}
Ultimately, the lowest energy electrons/positrons (resp. photons) in the shower start interacting with the medium mostly via ionisation (resp. Compton scattering) and the shower stops developing. The critical energy $\epsilon_c$ for which this happens is roughly defined as the energy for which the electron bremsstrahlung and ionisation rates are equal (in fact the energy at which the ionisation loss per $X_0$ is equal to the electron/positron energy). Denoting with $Z$ the atomic number of the medium, the critical energy can be approximated by~\cite{PDG}:
\begin{align}
\label{eq:criticalE}
\epsilon_c \sim \frac{610 ~\rm{  MeV}}{Z+1.24} \ .
\end{align}
A direct consequence of the two energy loss mechanisms described above is that one typically expects that in a thick target the differential density distribution should be dominated by electron/positrons around the critical energy. 

In order to describe more quantitatively the electromagnetic shower, we will broadly follow the approach of  Rossi and Griesen~\cite{RevModPhys.13.240} as reported in~\cite{Lipari:2008td}. We refer to Appendix~\ref{sec:app1} for details. The first step is to obtain the differential energy spectra of both electrons and positrons in the photon-induced electromagnetic shower. In this formalism $e^+$ and $e^-$ are treated on equal footing, neglecting initially 
the influence of ionisation or Compton scattering on the shower. This reads:
\begin{align}
\frac{dn^0_e (E, \Ez,t)}{dEdt} = \frac{1}{\Ez \sqrt{2 \pi} } \left[ \frac{G_{\gamma \to e} (s) }{\sqrt{\lambda_1''(s) t }} \left( \frac{E}{\Ez }\right)^{-1-s} e^{\lambda_1(s) t}\right] \; \; ,
\end{align} 
where the functions $\lambda_1(s)$ and $G_{\gamma \to e} (s)$ are defined in Appendix~\ref{sec:app1}, $s$ is the shower age defined in Eq.~\eqref{eq:showerage}, and $E_\gamma$ is the energy of the primary photon. The two auxiliary functions are constructed from the cross-sections for pair-production and bremstrahlung (along with several of their momenta) and thus include the details of the underlying physical processes leading to photons and positrons/electrons creation in the shower following the approach of Rossi and Griesen~\cite{RevModPhys.13.240}.

In a second step, an approximate solution including the cut-off effect from ionisation/Compton scattering can  be obtained by multiplying $\nez$ by a cut-off function $p_1$:
\begin{align}
\frac{d\ne (E, \Ez,t)}{dEdt} = \frac{d\nez (E, \Ez,t)}{dEdt} ~p_1 \!\! \left( s(\frac{\epsilon_c}{\Ez},t ) , \frac{E}{\epsilon_c} \right)  \,,
\end{align} 
and we will approximate the function $p_1$ by its value at the maximum of the shower ($s=1$)~\cite{Lipari:2008td}.

Finally,  the electrons and positrons differential track-length $\frac{dT_\pm}{dE}$ is obtained by integrating over the depth of the full shower and over the energy distribution of primary photons (with $E_{\rm ini}$ the energy of the primary proton initiating the shower). More precisely,
\begin{align}
\label{eq:tlanalytical}
\frac{dT_\pm}{dE} = \frac{1}{2} \int_0^{E_{\rm ini}} dE_\gamma  \int_0^\infty dt ~\frac{dn_e(E, E_\gamma,t)}{dEdt} ~\frac{d N_{\gamma} (E_\gamma)}{dE_\gamma} \ ,
\end{align}
where the factor 1/2 comes from the fact that the analytical approach does not distinguish between electrons and positrons. Note that $\frac{dT_\pm }{dE}$ has dimension of GeV$^{-1}$.
Intuitively, the quantity $\frac{dT_\pm}{dE}dE$ represents the total path length in the dump, in radiation length units, taken by $e^+/e^-$ with energy in the interval between $E$ and $E+dE$. This acts as an effective target length for LDM production, allowing the complicated EM shower to be condensed down to an effective fixed target experiment, as discussed in the following.

\begin{figure}
	\centering
	\subfloat[]{%
		\includegraphics[width=0.45\textwidth]{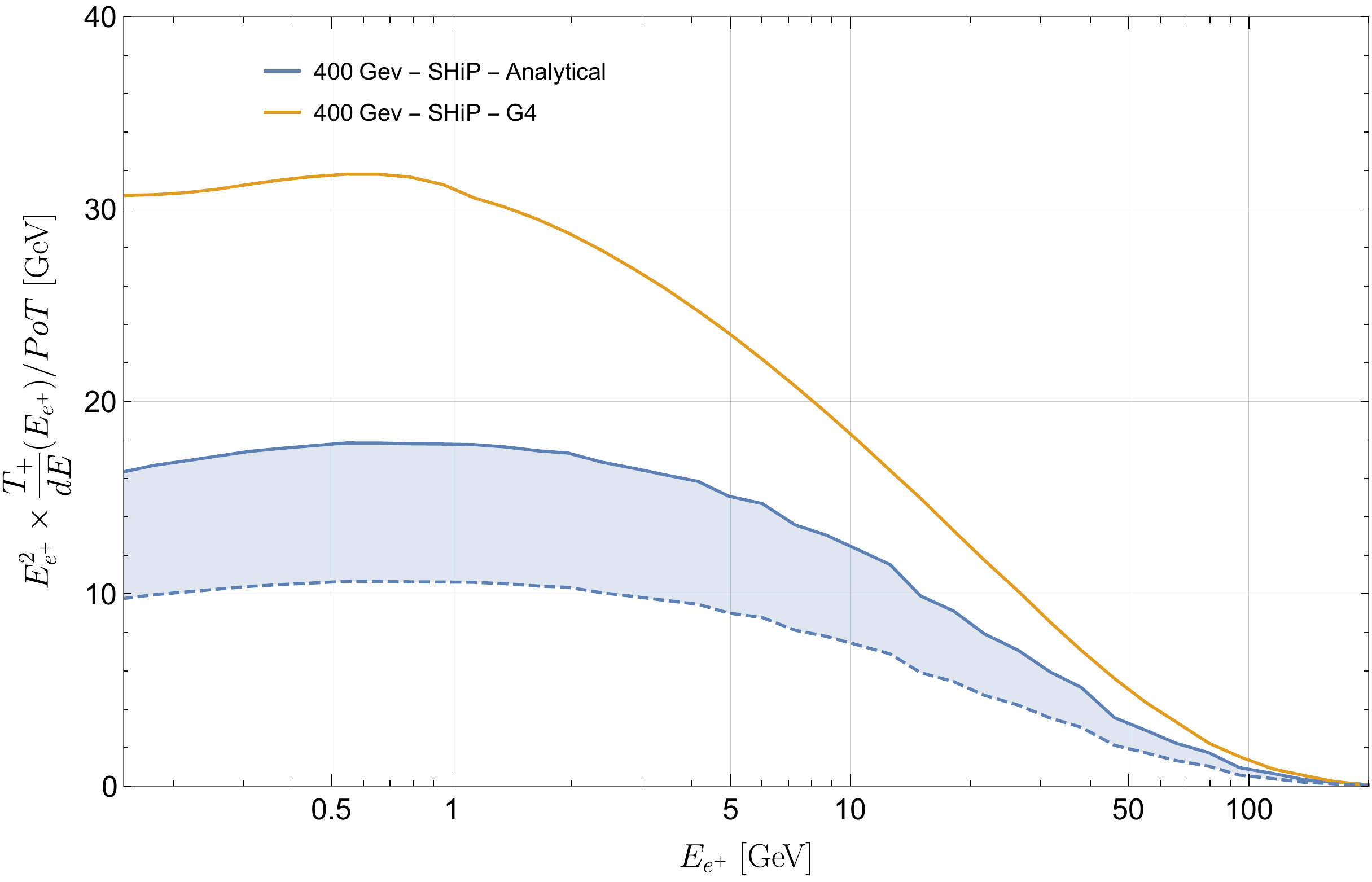}
	}\\
		\subfloat[]{%
		\includegraphics[width=0.45\textwidth]{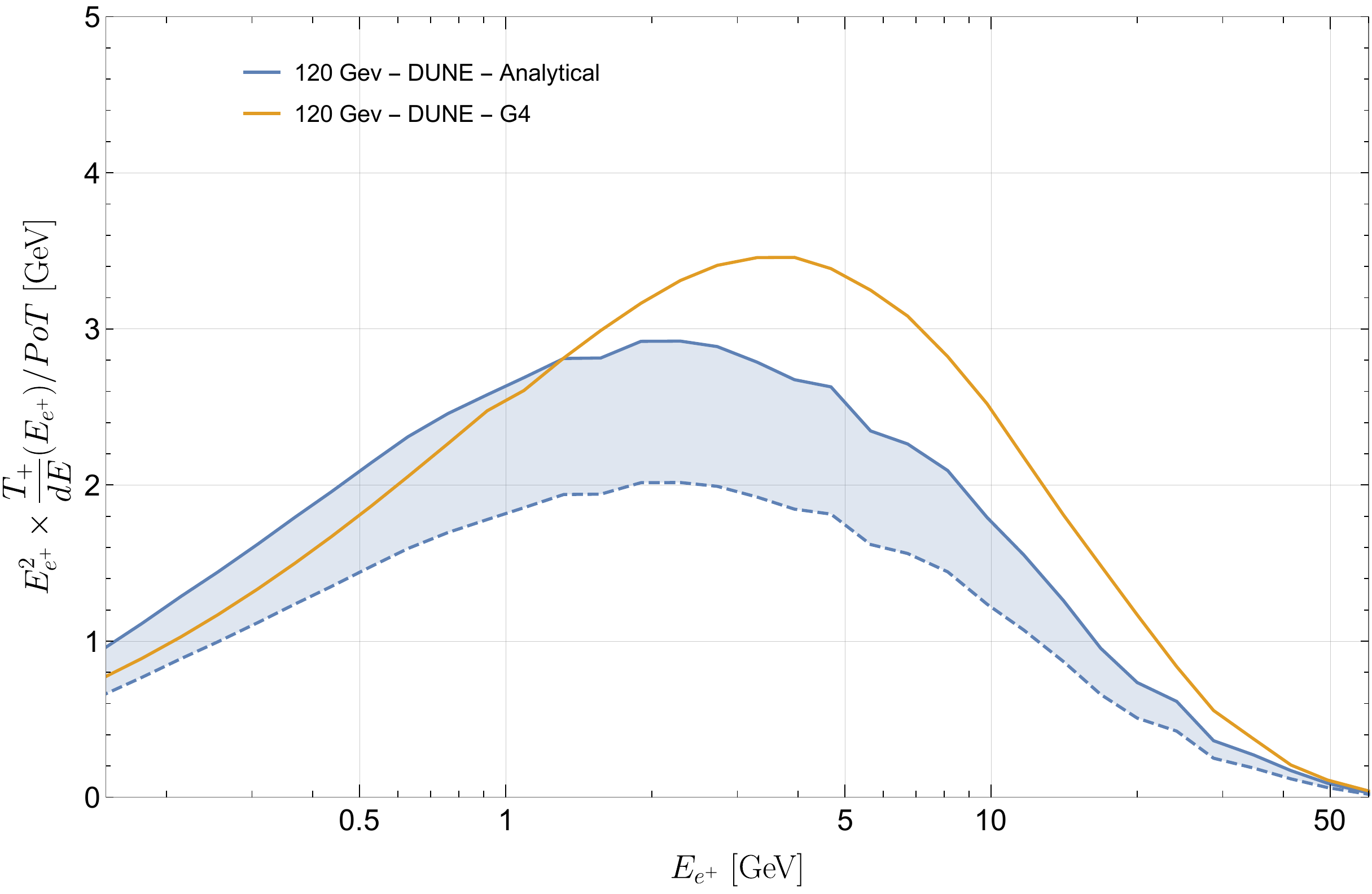}
	}\\%
		\subfloat[]{%
		\includegraphics[width=0.45\textwidth]{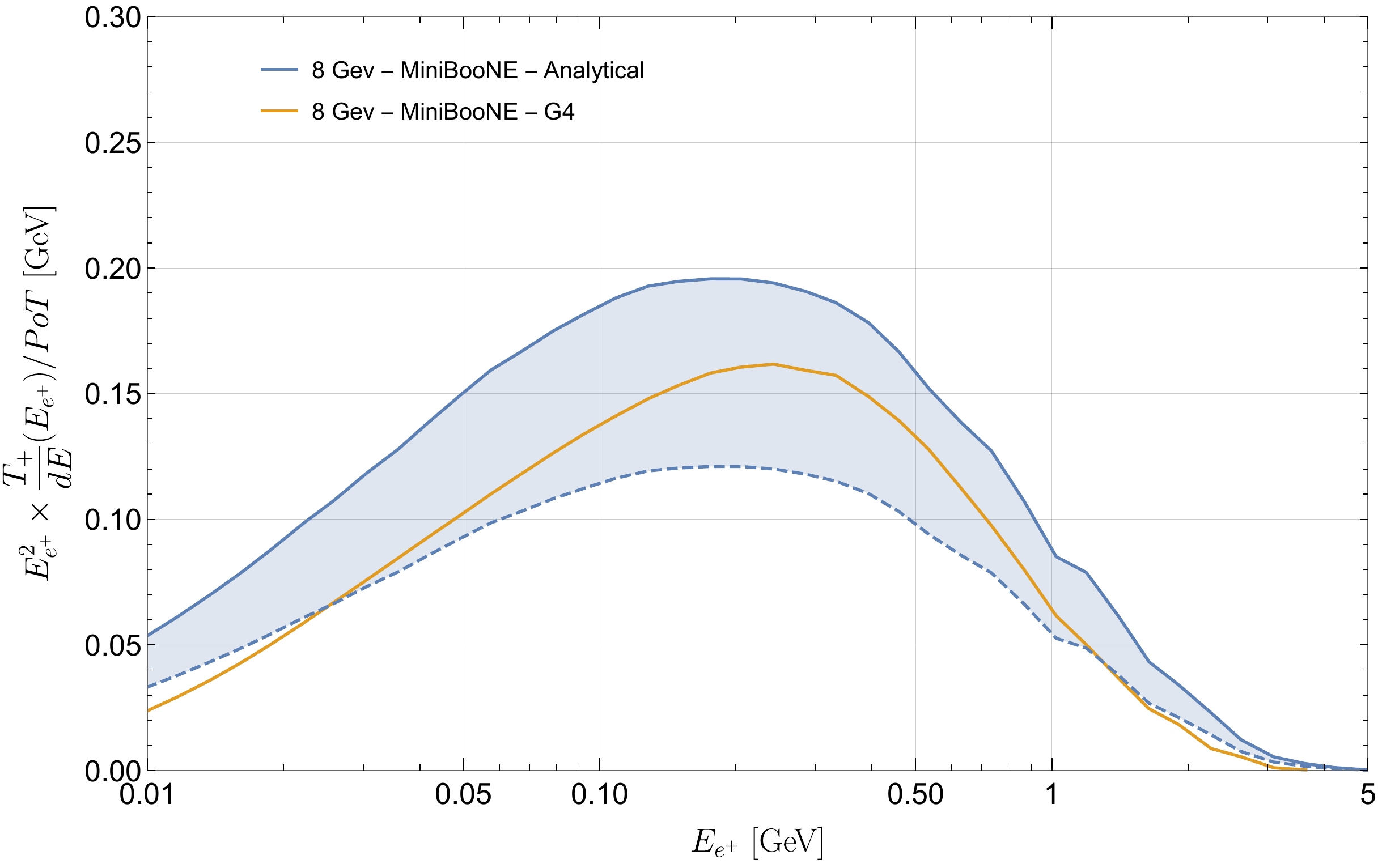}
	}%
	\caption{Differential track length times energy squared in GeV for the positrons in the showers generated in the SHiP, DUNE and MiniBooNE targets in arbitrary units. The yellow lines represent the results from  complete \texttt{\geant} simulations. The blue regions represent the results obtained from the semi-analytical approach described in the text, with the upper lines obtained by fixing  in Eq.~\eqref{eq:neutmesonyield} $L = \lambda_T$ (the nuclear interaction length) and the lower dashed lines corresponding to  $L = \lambda_c$ (the nuclear collision length) which is a more conservative choice. }
			\label{fig:FluxesW}
\end{figure}

We validate this approach in Fig.~\ref{fig:FluxesW}. In particular we show for reference the full result obtained from a \texttt{\geant} simulation~\cite{AGOSTINELLI2003250}, as is described in the next sections. We present the positrons track-length times energy squared distribution as function of the energy of the positrons. The semi-analytical approach carries an important uncertainty in that it does not account for the full dynamics of the hadronic shower, and it assumes instead that the initial proton interacts only once. Accordingly, the number of nuclei targets is set in Eq.\eqref{eq:neutmesonyield} by what is assumed to be the ``active'' part of the target. We can either set $L$ to the nuclear interaction length, or we can make the more conservative choice of setting $L$ to the nuclear collision length, thus ensuring that the incoming proton would not loose energy before generating the shower. The results obtained 
for these two choices delimit the  blue region in Fig.~\ref{fig:FluxesW}, 
which can be taken as a proxy for the typical uncertainty associated to the semi-analytical procedure. 
In any case, we find a very good agreement with the full numerical approach 
 for the experiments with the lower beam energies.  For the high-energy case of SHiP, we still obtain an acceptable agreement given the significant simplifications involved in the analytical approach which does not include the effects of secondaries, whose relevance increases with increasing beam energy.

We further observe that the semi-analytical approach becomes more conservative with increasing proton beam energy (in SHiP for instance) as secondary mesons carry enough energy to generate sizeable sub-showers of their own. Note that our final results will in any case be based on the complete numerical simulation shown in orange in Fig.~\ref{fig:FluxesW}.

\begin{figure}
 \centering
 \includegraphics[width=0.5\textwidth]{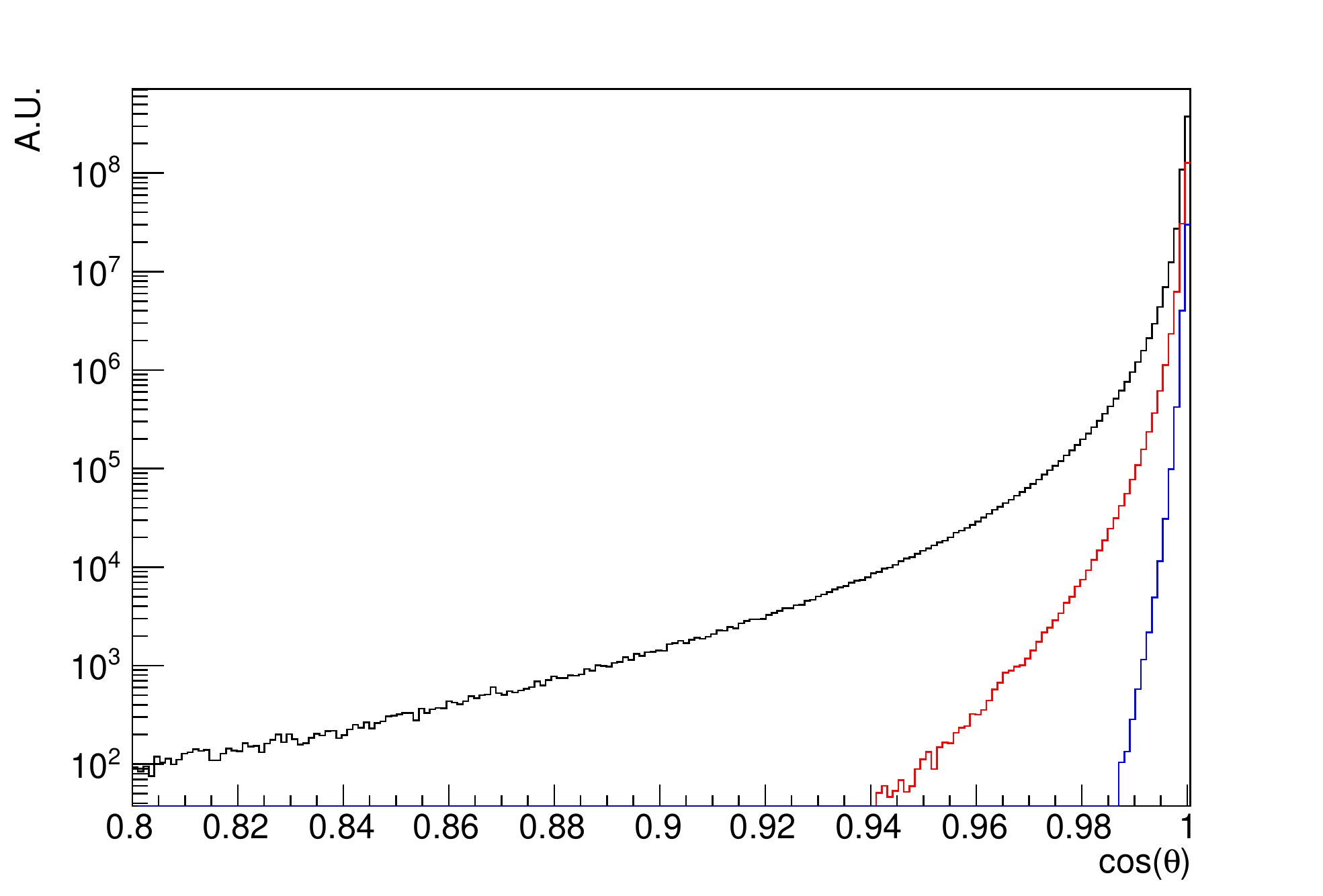}
 \caption{Angular distribution of positrons produced by the 120 GeV proton beam from the Fermilab accelerator in the DUNE target.
 The black, red and blue lines refer, respectively, to positrons with a 1 GeV, 3 GeV, and 8 GeV energy threshold. The normalisation of each curve is proportional to the total positron yield applying the corresponding energy threshold. The angular distribution of  electrons, not displayed, 
 features a similar behaviour. 
 }
 \label{fig:trckpos}
 \end{figure}

One important comment is that this approach does not incorporate the angular distribution of the produced electrons/positrons. As can be readily inferred from the relative low energy of the peak of the spectrum in Fig.~\ref{fig:FluxesW}, the electrons/positrons angular distribution has a non-negligible width.
 Depending on the  geometry of the experiment (detector size and detector-dump distance), this effect can be critical, since it affects the angular distribution of the  LDM particles produced, and thus the signal yield. In this work, we accounted for it by evaluating the double-differential track length $\frac{dT_\pm(E,\Omega)}{dEd\Omega}$. As an example, Fig.~\ref{fig:trckpos} shows the angular distribution of positrons produced in the DUNE target by the 120 GeV Fermilab proton beam.
 
\subsection{LDM production channels}
\label{sec:model}

\paragraph{LDM model building and Lagrangian }

The procedure described above is completely general and can be applied to any light new 
particle coupling to the electrons/positrons or to the light quarks (for instance axion-like particles and milli-charged particles).
For concreteness, in this work we focused on the case of a LDM scenario where sub-GeV DM particles 
interact with the SM  via a dark photon mediator $V^\mu$ (with 
field strength $F^{\prime\mu\nu}$ and dark gauge coupling $g_D$). The corresponding Lagrangian contains the following terms:
\begin{align}
\label{eq:Lag}
     \mathcal{L} \supset -\frac{1}{4}F^{\prime\mu\nu}F^{\prime}_{\mu\nu}
-\frac{1}{2}\frac{\varepsilon}{\cos\theta_w}B_{\mu\nu}F^{\prime\mu\nu}  - V^{\prime}_\mu g_D \mathcal{J}_{D}^{\mu} \ ,
\end{align}
where the parameter $\varepsilon$ weights the kinetic mixing, $B_{\mu\nu}$ the hypercharge field strength, and $\mathcal{J}_{D}^{\mu}$ is the dark gauge current, which depends on the details of the dark sector. After electroweak symmetry breaking, and 
after performing a standard  redefinition of the photon field $\gamma \to \gamma -\varepsilon V^\prime$ to diagonalise the kinetic term, the dark photon also acquires a $\varepsilon$-suppressed interaction with 
the SM electromagnetic current:
\begin{align}
     \mathcal{L} \supset - V^{\prime}_\mu e \varepsilon \mathcal{J}_{\rm{em}}^{\mu} \ .
\end{align}

Note that the dark photon mass $m_V$ can originate either from the Stueckelberg mechanism or from the VEV of a dark Higgs boson. The latter typically constitutes an important part of the phenomenology if it has the same mass as the dark matter candidate~\cite{Choi:2016tkj,Darme:2017glc,Darme:2018jmx}; on the contrary, it can basically decouple if it is heavier than the dark photon. Here we will consider explicitly the second scenario.
Finally, specifying the precise nature of the dark matter candidate $\chi$ is not critical for the scope of this work. 
In order to compare our result with the recent limits from the MiniBooNE collaboration, we considered a complex scalar dark matter candidate, although our conclusions also apply for other standard choices (Majorana dark matter, pseudo-Dirac dark matter with a small mass splitting, etc...) since their production and detection mechanisms are similar. For the case of a complex scalar the dark current is given by:
\begin{align}
    \label{eq:darkcurrent}
    \mathcal{J}_{D}^{\mu} = i \left( \chi^* \partial^\mu \chi - \chi \partial^\mu \chi^* \right) \ .
\end{align}
As long as $m_V > 2 m_\chi$, the interaction in Eq.~\eqref{eq:Lag} leads to rapid dark photon decay into dark matter particles: this is the so-called \textit{invisible decay scenario} on which we focus.
Note that often in the literature 
an extra factor of $1/2$ is included in the normalisation of the dark gauge current  in Eq.~\eqref{eq:darkcurrent}. 
Thus, when relevant to carry out proper comparisons, we have rescaled the existing limits on 
the dark gauge coupling in Eq.~\eqref{eq:Lag} to account for the choice of normalisation.\footnote{Most notably,
 the recent works using the convention with an extra factor $1/2$ include the prospects for the SHiP collaboration as reported in, e.g~\cite{Buonocore:2018xjk,Ahdida:2654870}, as well as the study of the projected sensitivity 
of the NO$\nu$A near detector in~\cite{deNiverville:2018dbu}.}

\paragraph{Main production channels and cross-sections}

For low mass dark sectors, the main production mechanisms for dark photon from the hadronic development of the shower are from the decay of light unflavored mesons. Depending on the mass of the dark photon, the dominant meson decay process are  $\pi^0 \rightarrow \gamma V$, $\eta, \eta' \rightarrow \gamma V$ or $\rho, \omega \rightarrow V \rightarrow \chi \chi^*$ (in case the dark photon decays into dark sector particles). The typical branching ratio is given by
\begin{align}
\mathcal{BR}(\pi^0\rightarrow V \gamma) = 2 \varepsilon^2 \left( 1- \frac{m_V^2}{M_{\pi^0}^2}\right)^3 \ ,
\end{align}
In particular, note that there is no $\aem$ insertion so that this process is only mildly suppressed.

\begin{figure}
	\subfloat[]{%
		\includegraphics[width=0.24\textwidth]{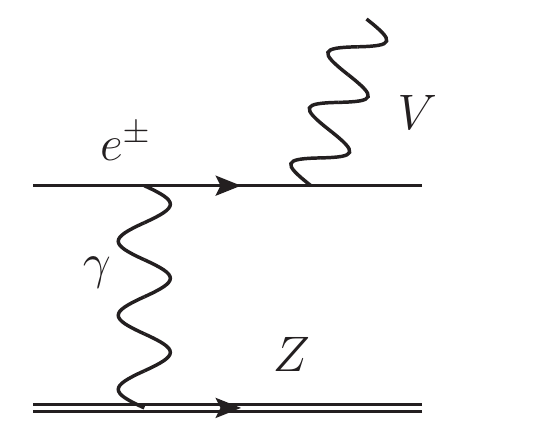}
		\label{fig:diagbrem}
	}%
	\hspace{0.025\textwidth}
		\subfloat[]{%
		\includegraphics[width=0.2\textwidth]{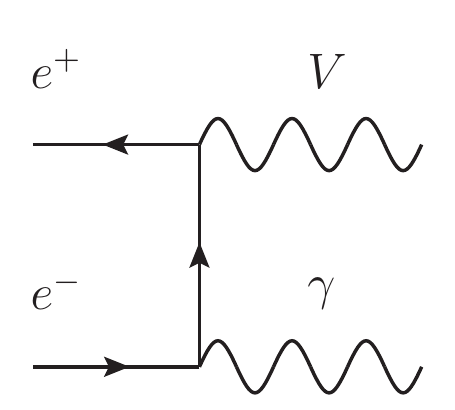}
		\label{fig:diagAssoc}
	}%
		\hspace{0.025\textwidth}
		\subfloat[]{%
		\includegraphics[width=0.2\textwidth]{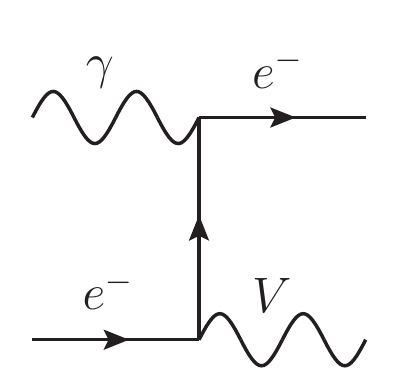}
		\label{fig:diaginvCompton}
	}%
		\hspace{0.025\textwidth}
		\subfloat[]{%
		\includegraphics[width=0.2\textwidth]{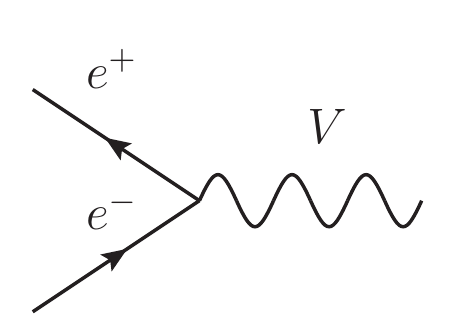}
		\label{fig:diagResonant}
	}%
	
	\caption{Dominant processes for dark photon production during the electromagnetic development of a shower}
	\label{fig:diags}		
\end{figure}

On the other hand,  hadronic showers develop a large electromagnetic component from the radiative decays of light neutral mesons $\pi^0, \eta$. All relevant processes here depend on the density of the relevant targets (either nuclei for bremsstrahlung or atomic electrons for positron/photon processes). While the dominant production mechanism for vector mediators in electron beam dumps is mostly via electron bremsstrahlung, it was recently realised that production mechanisms based on secondary positrons  can dominate in the low mass ranges~\cite{Marsicano:2018glj}.
Since in hadronic showers the yield of secondary electrons and positrons is almost the same, positron-related processes dominate the LDM production rate.

Denoting $E_\pm$ the energy of the incoming positron/electron in the lab frame, the main processes responsible for dark photon production by secondary $e^+/e^-$, illustrated in Figure~\ref{fig:diags}, are the following:
\begin{itemize}
\item Bremsstrahlung of electrons and positrons off  nuclei, $e^\pm N \rightarrow e^\pm N V$ with typical cross section:
\begin{align}
\sigma_{\rm brem} \simeq \frac{ 4 \varepsilon^2 \aem^3 }{3}\sqrt{1-\frac{m_V^2}{E_\pm^2} } \frac{\xi}{m_V^2} \log \left( \frac{1}{\rm{Max} (\frac{m_e^2}{m_V^2},\frac{m_V^2}{E_\pm^2})}\right) \ ,
\end{align}
where $\xi \sim Z^2$ is the effective flux of photon from the accelerated nuclei in the incoming electron/positron frame~\cite{Bjorken:2009mm}. We observe that $\sigma_{\rm brem}$ increases quadratically for small dark photon mass, but is, however, severely suppressed by $ \aem^3$. Also, the emitted dark photons are
typically very energetic, since they carry most of the energy of the initial $e^+/e^-$, with the median value for $E_V$ given by:
\begin{align}
 \langle E_V \rangle = E_\pm \left( 1  - \rm{Max} (\frac{m_e^2}{m_V^2},\frac{m_V^2}{E_\pm}) \right) \ .
\end{align}
This mechanism dominates the dark photon production in electron beam-dump experiments, due to the fact that it is enhanced for very energetic primary electrons (see the comparison with the resonant production mode in~\cite{Marsicano:2018glj,Marsicano:2018krp}). In the proton-shower induced environment, both the electrons and the positrons are secondary particles, and therefore they contribute equally to the bremsstrahlung production rate. We review in more detail this production mechanism and our numerical approach for this process in Appendix~\ref{sec:app2}.
\item Direct  positrons annihilation on target atomic electrons $e^+ e^- \rightarrow V \rightarrow \chi^* \chi$~\cite{Nardi:2018cxi}. This process can be divided in two main regimes: resonant and off-shell. 
In the resonant regime, the dark photon is produced on-shell and the cross section is given by:
\begin{align}
\sigma_{\rm res} = \frac{2 \pi^2\varepsilon^2 \aem }{m_e} \delta ( E_+-\frac{m_V^2}{2m_e}) \ .
\end{align}
While this process can only occurs around the resonant energy (depending on the width of the 
dark photon, which is here relatively large due to the dark decay $V \to \chi \chi^*$), it is 
still important because it is only suppressed by $\aem$~\cite{Nardi:2018cxi}. Furthermore, given the restricted kinematics, the energies of the incoming positron and of the outgoing dark photon are related by
\begin{align}
E^{\rm res}_V = \frac{m_V^2}{2m_e} \ .
\label{eq:resE}
\end{align}
This allows to estimate the range of the accessible dark photon masses as:
\begin{align}
\label{eq:mth}
    m_V^{th} \gtrsim \sqrt{2 m_e E_{\rm th}} \ ,
\end{align}
where $E_{\rm th}$ is the experimental detection threshold. Note that the above expression is very conservative in that it assumes that the dark photon energy is entirely transmitted to the detector.

In the off-shell regime, $\chi^*\chi$ pairs are produced via exchange of an off-shell $V$,
and this process can be relevant especially when considering a large dark gauge coupling $\alpha_D \sim 0.1$. Accounting for off-shell  $\chi^*\chi$ production requires including in the resonant positrons annihilation the finite $V$ width, and considering the full four-particle $s$-channel reaction $e^+ e^- \to V^* \to \chi^* \chi$. 
More precisely, in the limit where the center-of-mass (CM) energy $\sqrt{s}$ is much larger than $m_V$ (particularly relevant for small, MeV-scale dark photon), the off-shell contribution can be estimated as:
\begin{align}
\label{eq:prodoffshell}
\sigma_{\rm off-shell} = \frac{\pi \aem \varepsilon^2 \alpha_D}{6 m_e E_+} \ .
\end{align}

\item Associated production from positrons in the shower, $e^+ e^- \rightarrow \gamma V$. In the limit where $E_+ m_e \gg m_V^2$, the cross section becomes
\begin{align}
\label{eq:assocprod}
\sigma_{\rm assoc} \simeq \frac{2 \pi \varepsilon^2 \aem^2 }{m_e E_+} \log (\frac{2 E_+}{m_e}) \ ,
\end{align}
which is typically $\aem^2$ suppressed but is enhanced by a $1/m_e$ factor. Note that, due to the presence of an additional photon in the final state, in this case the energy of the emitted dark photon can  differ from $\frac{m_V^2}{2m_e}$. In fact, as shown in Appendix~\ref{sec:app3}, around half of the dark photons from associated production retain most of the energy of the incoming positron $E_V \sim E_+$. 
Compared with the direct $e^+ e^- \to V^* \to \chi^* \chi$ off-shell regime in Eq.~\eqref{eq:prodoffshell}, the log-enhanced $\aem \log (2 E_+/m_e)$ term is replaced by the dark gauge coupling term $\alpha_D$. Therefore, in case  $\alpha_D \gtrsim 0.1$, $\sigma_{\rm assoc}$ is negligible with respect to $\sigma_{\rm off-shell}$.
On the other hand, the experimental energy threshold tends to suppress both these processes with respect to  bremsstrahlung. 
\end{itemize}
Note that for the processes leading to an on-shell dark photon, $V$ decays to a $\chi^*\chi$ pair with near $100 \%$ branching ratio for sizeable dark gauge coupling, thus allowing to easily derive the LDM production yield.

Finally, the electromagnetic shower further contain a significant number of photons, making the Compton-like scattering process $\gamma e^- \rightarrow e^- V$ also a potentially relevant production channel. In the limit where $E_+ m_e \gg m_V^2$, the cross section becomes
\begin{align}
\sigma_{\rm Compton} \simeq  \frac{\sigma_{\rm assoc} }{2} \simeq  \frac{\pi \varepsilon^2 \aem^2 }{m_e E_+} \log (\frac{2 E_+}{m_e}) \ ,
\end{align}
which is also $\aem^2$ suppressed. Note that this cross section falls much faster than that 
for associated production at larger dark photon mass. We present a thorough description of the impact of this channel, comparing it to the associated and bremsstrahlung production channels, in Appendix~\ref{sec:app3}.

\paragraph{Comparison and total production rates}

\begin{figure}
	\centering
		\includegraphics[width=0.48\textwidth]{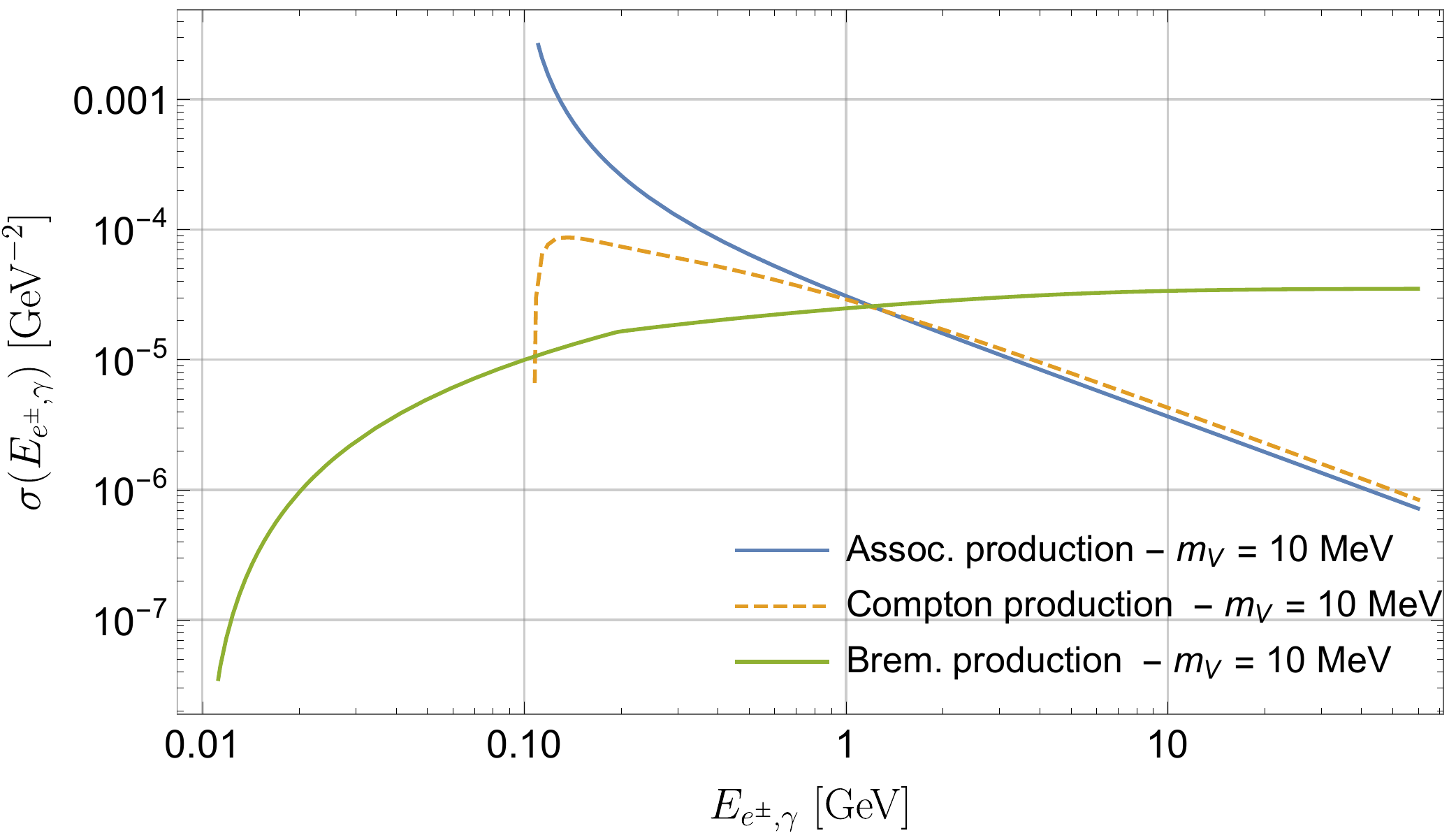}
	\caption{Production cross section for the associated $e^+ e^- \to \gamma V$ and Compton-like process $\gamma e^- \to e^- V$ as function of the energy of the incoming particle (either a photon $E_\gamma$, a positron or an electron with energy $E_{e^{\pm}}$) in the laboratory frame. We chose $\varepsilon = 0.001, m_V = 10$ MeV.}
			\label{fig:allCSofE}
\end{figure}

In all the experiments we have considered the associated production process is often sub-dominant compared to the bremsstrahlung or to the resonant production mechanism. Indeed, the latter strongly dominates due to its $\epsilon^2 \alpha_{\rm em}$ scaling when enough positrons with adequate energy $\sqrt{m_V^2/ (2m_e)}$ are produced in the showers.  On the other hand, as can be seen in Figure~\ref{fig:allCSofE}, the bremsstrahlung cross section saturates at high incoming energy, while both associated and Compton-like process decrease due to their $1/s$ dependence. This implies that, even for very small dark photon masses where  there is a $1/m_V^2$ enhancement, bremsstrahlung production gets contributions from positrons in the full range of energies available in the shower. 
Furthermore, in the opposite limit of large dark photon masses, where the resonant dark photon energy $E^{\rm res}_V$, see Eq.~\eqref{eq:resE}, is larger than the beam energy
and resonant production cannot occur, both associated and Compton-like processes are also forbidden. In this case, the bremsstrahlung process has access to a larger CM energy since it corresponds to an interaction with the nucleus,  and can be effective up to $E_{\pm} \sim m_V$.

\begin{figure}	
\includegraphics[width=0.45\textwidth]{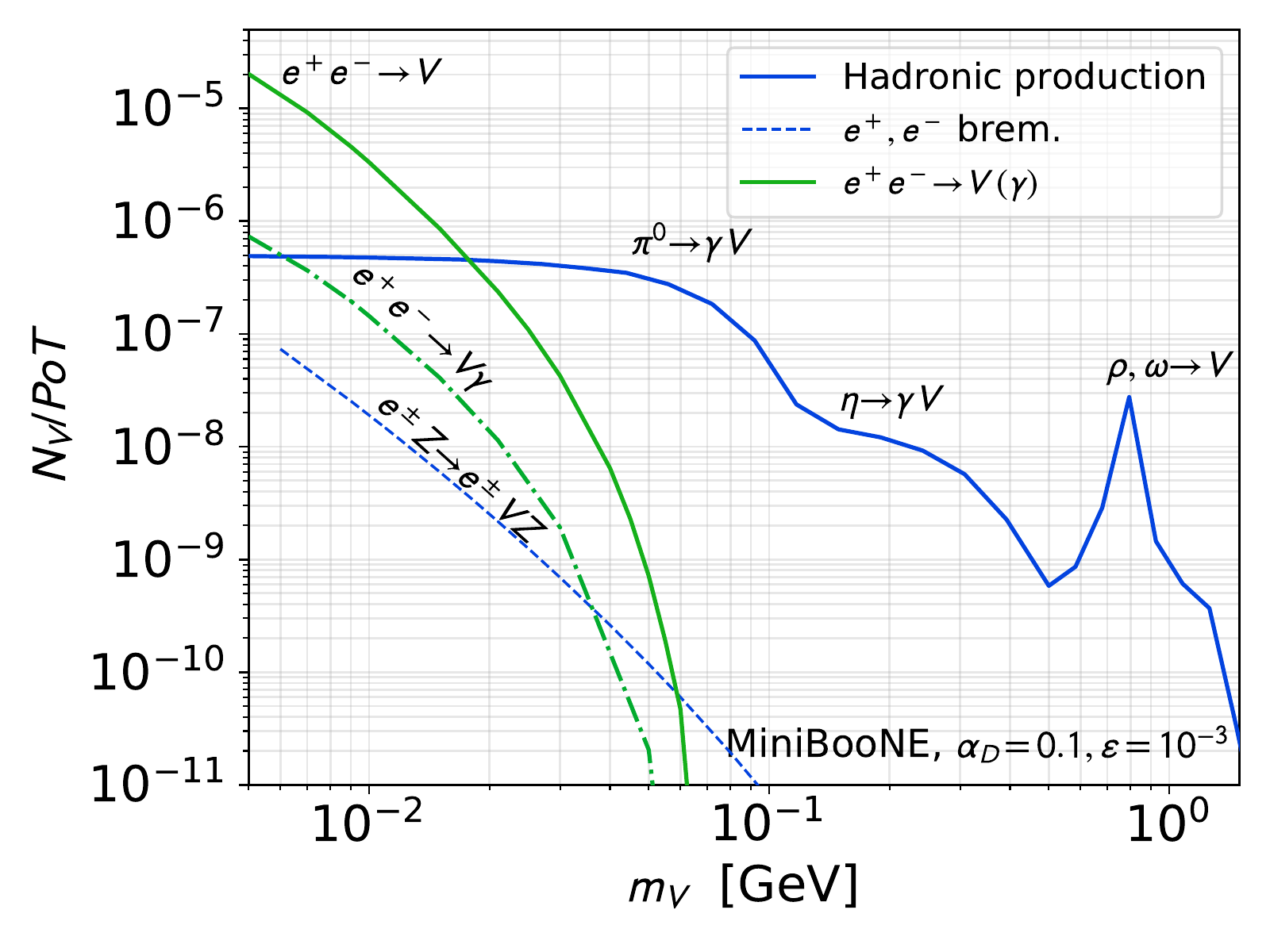}
\caption{Dark photon production rate per proton-on-target for the MiniBooNE experiment as function of the dark photon mass $m_V$.
The  shower-induced leptonic production processes are shown in green:
electron/positron bremsstrahlung (dashed line) and resonant $e^+ e^- \to V, V\to \chi \chi^*$ (solid line), 
The blue line corresponds to the rate for standard hadronic production processes. We have applied basic cuts on the \texttt{\geant} objects:  their angle $\theta$ with respect to the beam axis is selected such that $\sin \theta < 0.2$, and their kinetic energy should be larger than $10$ MeV. }
\label{fig:NV}
\end{figure}

In order to illustrate the respective importance of mesons decay process with respect to shower-induced ones, we present in Fig.~\ref{fig:NV} the corresponding dark photon production rates for the $8$ GeV proton beam servicing the MiniBooNE experiment. We used the full \texttt{\geant} simulation described in the next section to obtain both the distribution of light mesons and the track length of secondary positrons. 
Interestingly, the secondary production strongly dominates in the lower mass regimes. This is both due to the fact that the meson production saturates in this regime and that the showers provide an abundant number of positrons and electrons with enough energy to produce such light dark photons. Both hadronic and shower-based processes have the same production rate for a dark photon mass  around $m_{\rm cross} \sim 16$ MeV. This ``crossing'' mass depends more generally on the energy available in the initial proton beam as well as on the material of the target. For instance, for the $120$ GeV beam from Fermilab's main injector, which will be used by the DUNE experiment, $m_{\rm cross} \sim 20$ MeV, while for the proposed SHiP experiment with access to the SPS $400$ GeV beam and a high-Z material target, $m_{\rm cross} \sim 30$ MeV.

\begin{figure*}[t]
\centering
\includegraphics[width=.8\textwidth]{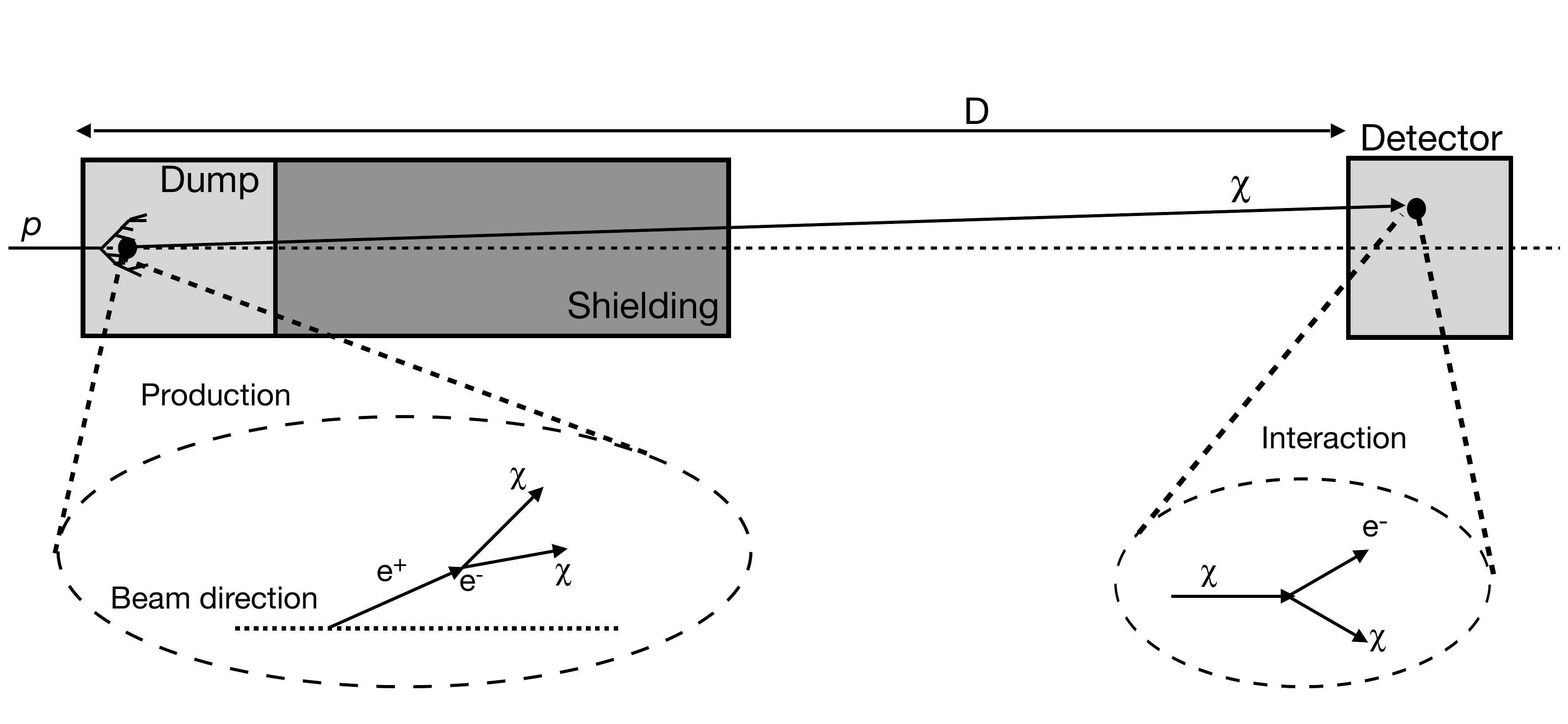}
\caption{\label{fig:scheme}
Typical setup of a proton beam-dump experiment. The proton beam impinges on a thick target, where LDM particles are produced by secondaries - the inset shows the production of a LDM particle pairs from $e^+e^-$ annihilation. LDM particles then propagate straight toward a downstream detector at distance $D$, where they are 
revealed via the scattering on atomic electrons and nuclei.}
\end{figure*}

 \subsection{Experimental LDM production and detection}

The typical setup of a proton beam-dump experiment is shown in Fig.~\ref{fig:scheme}. The primary proton beam impinges on a thick target, where LDM particles are produced. These propagate straight towards a detector with cross size $S$ placed at distance $D$ downstream that reveals them. A sizeable amount of shielding material is placed between the dump and the detector to range out all other particles produced by the primary beam, except neutrinos.

In all the experimental setups considered in this paper, the distance between the target and the detector is much larger than the length of the target, so that the entire shower can be approximated as starting from the initial vertex. Therefore, the number of LDM particles emitted through a process characterised by a cross section $\sigma$ can be computed as:
\begin{align}
\mathcal{N} = \frac{\mathcal{N}_A X_0 \rho}{A} \int_0^{E_{ini}}  dE~\frac{dT_\pm(E)}{dE}~\sigma (E) \ ,
\end{align}
where the $X_0$ is the radiation length of the material, $\rho$ its mass density, $A$ its atomic mass and $\mathcal{N}_A  = 6.022 \times 10^{23}$.\footnote{Note that the cross section has to be expressed in cm$^{-2}$.} Depending on the production process being considered, $\frac{dT_-}{dE}$ and/or $\frac{dT_+}{dE}$ should be used. Similarly, the differential yield 
 $\frac{dN}{dE_\chi}$ can be obtained by replacing $\sigma~\rightarrow~\frac{d\sigma}{dE_\chi}$. 
 
As discussed before, this approach does not incorporate  the angular distribution of the produced electrons/positrons. A rough estimate of the detector geometric acceptance is $\varepsilon_T \sim S/(\theta_\chi D)^2$, with $\theta_\chi$ being the average LDM emission angle. 
This has to be computed by convolving the different processes that are ultimately resulting to LDM production in the thick target: the production of primary neutral mesons in the hadronic shower, their decay to photons, the development of the EM shower, and the LDM production by electrons and positrons. Furthermore, the angular shape of each of these processes has its own energy dependency. Therefore, a numerical approach is here unavoidable.

If the $\chi$ couples diagonally with the $V$, the two main processes responsible for the interaction with the detector are the elastic scattering off electrons and the quasi-elastic scattering off nucleons. In the electron case, since  $m_e \ll m_{V}$, the electron carries most of the impinging $\chi$
energy and gives rise to an electromagnetic shower in the detector. In the nucleon case, instead, due to the nucleon larger mass, the recoil energy is typically lower, making the signal corresponding to this process more difficult to
identify. For this reason, in this work we focus on the $\chi - e$ scattering process only. The differential cross section for $\chi e \rightarrow \chi e$ scattering  with respect to the electron recoil energy $E_f$ in the laboratory frame is~\cite{Batell:2014mga}:  
\begin{align}
\label{eq:csscat}
\frac{d \sigma_{f,s}}{d E_f} =  4 \pi \varepsilon^2 \alpha \alpha_D \frac{2 m_e E^2 - f_{f,s}(E_f)(E_f - m_e)}{(E^2 - m_\chi^2)(m_{V}^2 + 2 m_e E_f - 2 m_e^2)^2}    
\end{align}
where $E$ is the incoming $\chi$ energy and $f$ and $s$ stand for fermion and scalar $\chi$ respectively; $f_f(E_f) = 2 m_e E -m_e E_f + m_\chi^2 + 2 m_e^2$, $f_s(E_f) = 2 m_e E + m_\chi^2$. The total signal yield can then be obtained analytically, convolving the differential cross section with the incoming LDM distribution and the cut efficiency for electron recoil detection. 
Note that while in this paper we consider the detection of LDM via its scattering in the detector, the main idea of secondary dark photon production is relevant also for other types of dark sector searches.

We finally observe that, while in this work we focused on the case of LDM detection through the elastic scattering on atomic electrons, our idea also applies to LDM models predicting similar interaction mechanisms in the detector. For example, in inelastic dark matter scenarios (iDM)~\cite{TuckerSmith:2001hy}, if the splitting between the two dark $\chi_1 e^+ e^-$ states is small with respect to the beam energy scale, the leptons-induced LDM yield in the beam dump would not change significantly. At the same time, provided $m_{\chi_2} > m_{\chi_1} + 2m_e$, the expected signature in the detector would be either the direct decay $\chi_2 \rightarrow \chi_1 e^+ e^-$ within the detector when the $\chi_2$ state is sufficiently long-lived, or the non-diagonal scattering $\chi_1 N \rightarrow \chi_2 N$, with $N$ an atomic nucleus, followed by the decay $\chi_2 \rightarrow \chi_1 e^+ e^-$. In  both cases, the result is a significant energy deposition in the detector. In particular, we note that in the limit where the heavy state $\chi_2 $ has a decay length much larger than the distance to the detector, the lower boost factor of secondary production events will enhance the detection prospects. We will investigate the effect of shower-induced iDM production in a future work (see e.g.~\cite{Izaguirre:2017bqb,Darme:2017glc,Berlin:2018jbm,Berlin:2018pwi,Darme:2018jmx,Mohlabeng:2019vrz,Tsai:2019mtm,Jodlowski:2019ycu} for recent works discussing the iDM physics case).

\section{Numerical evaluation}
\label{sec:numerics}
To re-evaluate  
the exclusion limits implied by existing proton beam-dump results when  
the lepton-induced secondary production processes are properly included,
and to estimate the sensitivity of planned experiments, 
the expected number of signal events within the detector has to be computed as a function of the model parameters, and compared with the background yield. We performed the calculation of the signal yield numerically, decoupling the evaluation of the LDM production in the beam-dump from the subsequent propagation and detector interaction as described below.

All the necessary numerical ingredients, including in particular the track length distributions used to describe the electrons and positrons from the sub-showers are available on \href{https://zenodo.org/record/3890984}{the Zenodo online repository}~\cite{celentano_andrea_2020_3890984}.

\subsection{LDM production}

The evaluation of the LDM production in the dump was further factorized into two independent steps: i) the calculation of the electrons and positrons track-length in the target, and ii) the computation of the LDM differential yield from $e^+$ interactions.

For each of the detector setups that we have considered in this work, and that are described in the next section, we have computed $\frac{T_\pm(E,\Omega)}{dEd\Omega}$, i.e. the electrons/positrons differential track-length distribution as a function of the particle energy and angle, by means of 
a \texttt{\geant} simulation. 
We have used the standard \texttt{G4EmStandardPhysics} physics list to describe EM interactions, and the \texttt{FTFP\_BERT\_HP} physics list to parameterize hadronic reactions.
We have developed a custom class, inheriting from \texttt{G4SteppingAction}, that records, for each electron and positron step in the target, the corresponding particle energy and direction. The output of the simulation is the distribution $\frac{T_\pm(E,\Omega)}{dEd\Omega}$ for discrete bins of the two observables. For the energy, we have used a bin width $\Delta E$ corresponding to $\sim 0.1\%$ of the primary proton beam energy. Since in the simulation, with default physics lists settings, the typical energy loss for each positron step inside the dump volume is already much smaller than $\Delta E$, we did not include any explicit step limiter. Finally, to speed-up the calculation, we introduced for all particles an energy threshold equivalent to the detection threshold, discarding from the simulation all particles falling below this value.
In order to make a fair comparison between the electron- and positron-induced production mechanisms with the ``traditional'' processes usually considered for  proton beam-dump experiments involving 
neutral mesons decays, in the simulations we have also sampled the differential distribution $\frac{dn_{M^0}(E,\Omega)}{dEd\Omega}$
 for $ M^0 =\pi^0,\, \eta$.

The LDM yield in the target was then computed using the \texttt{MADDUMP} software~\cite{Buonocore:2018xjk} and a modified version of the Monte Carlo generator \texttt{BdNMC}~\cite{deNiverville:2016rqh} depending on the production process. The former is a plugin for the \texttt{MadGraph5\_aMC}$@$\texttt{NLO} program~\cite{Alwall:2011uj,Alwall:2014hca} that allows to compute the differential yield of LDM particles in the target from the knowledge of $\frac{dT_\pm(E,\Omega)}{dEd\Omega}$. In particular, we used \texttt{MADDUMP} to generate a list of outgoing dark matter momenta from all the leptonic production channels, including the $s$-channel $e^+ e^- \to V^{(*)} \to \chi \chi^* $, the associated production and the bremsstrahlung processes. For the latter, we adopted the nuclear form-factor parameterisation described in Ref.~\cite{Bjorken:2009mm}. 
On the other hand the hadronic production processes were handled  by \texttt{BdNMC}. For the production via light meson decays, we used the light neutral meson distributions including secondary mesons as given by \texttt{\geant} (instead of the build-in empirical distributions) and we have simulated their decay to dark matter via the vector portal. For completeness, we have  further included the proton bremsstrahlung process and 
 dark photon production via resonant vector meson mixing as it is implemented in \texttt{BdNMC}~\cite{deNiverville:2016rqh} (in particular, the timelike form factor used in the production rate is derived from~\cite{Faessler:2009tn} and hence incorporates the effect of $\rho/\omega$ meson production).
We observe that, in the current version, both \texttt{MADDUMP} and \texttt{BdNMC} assume that all LDM particles are produced at the beginning of the target, neglecting the development of the EM shower in the corresponding volume. However, as already mentioned, this approximation is well justified by the much larger distance between the target and the detector.

The advantage of this dual approach, rather than handling together the description of the EM shower development and the production of LDM in a single simulation, is the fact that, for each considered experiment, the differential track length and the neutral mesons distribution have to be computed only once, thus saving a significant amount of computation time. Only the evaluation of the LDM yield has to be repeated for different values of $m_V$ and $m_\chi$. 

Finally, to account for the different materials in the target geometry, the procedure we adopted was to compute separately for each of them  the $e^+/e^-$ differential track length and the LDM yield using the procedure described before, summing the obtained results. To speed-up the calculation, only the materials with a non-negligible track-length relative weight ($\gtrsim 1\%$) were further considered.

\subsection{Detector interaction and normalisation}

We have used \texttt{BdNMC} to simulate the propagation and interaction of light dark matter with the detector. More precisely, we propagated the LDM  particles to the detector and estimated their intersection with the detectors using the internal \texttt{BdNMC} routines. The scattering probability as a function of the dark matter nature (complex scalar or Dirac-fermion) was estimated using Eq.~(\ref{eq:csscat}) (note that the complex scalar case was already present in the original \texttt{BdNMC} code). In order to simulate accurately the detector response, we added at the generator-level the selection cuts from the experiments. To speed-up the calculation, basic energy cuts were included directly in the cross section evaluation, while the more advanced ones (such as that on $E_e \theta_e^2$) were applied after the scattering events had been simulated.

Finally, starting from the knowledge of the sensitivity of a given experiment in terms of signal yield, the corresponding reach curve was sampled as follows. To reduce the number of free parameters, we adopted the standard choice $m_V=3m_\chi$ and $\alpha_D=0.1$. 
Observing that in the scenario considered in this work all LDM particles are produced promptly in the beam dump, we can expect that for a given set of reduced model parameters  the foreseen signal yield in the detector will  scale as:
\begin{equation}
    N_S(m_\chi,\varepsilon) = N^0_S(m_\chi) \cdot 
    \left( 
    \frac{\varepsilon}{\varepsilon_0}\right)^4  \; \; ,
\end{equation}
where $N^0_S(m_\chi)$ is the signal yield corresponding to the kinetic mixing parameter $\varepsilon_0$. 
 We can thus  obtain the limit for $\varepsilon$ by inverting the previous relation.

\section{Applications and examples}
\label{sec:expresults}

\begin{table*}[t]
	\centering
	\resizebox{1.0\textwidth}{!}{\begin{minipage}{1.\textwidth}
		\begin{tabular}{|c||c|c|c| c| c|c  | c| }
		\hline
				\rule{0pt}{14pt}Experiment & $E_{\rm beam}$&  Target & PoT  & D (m) & L/S (m/m$^2$) &  $E_{\rm cut}$ (scat)  & NoE $90\%$ \\
				\hline
				\hline
				\rule{0pt}{14pt}
			 MiniBooNE~\cite{Aguilar-Arevalo:2018wea} & $8$ GeV& Steel & $1.86 \cdot 10^{20}$  &$490$ &$12$ / $36$ & 75 MeV  & $2.3$ \\
			 \hline
 		    NO$\nu$A~\cite{Aguilar-Arevalo:2018wea} & $120$ GeV& C & $2.97 \cdot 10^{20}$  &$990$ &$12.67$ / $3.9
 		    \times 3.9$ & 500 MeV & $16.4$  \\
				\hline
				\rule{0pt}{14pt}
	    	SHiP~\cite{Anelli:2015pba} & $400$ GeV& W / Mo / Fe & $2 \cdot 10^{20}$  &$38$ & $3.2$ / $0.75\times3.2$ & $1$ GeV & $38$ \\
            \hline
            DUNE-PRISM \cite{DeRomeri:2019kic} & $120$ GeV& C & $7.7 \cdot 10^{21}$   &$574$ &$5$ / $3\times4$ & 50 MeV & $350$ ($54$) \\
						\hline
			\end{tabular}
			\end{minipage}}
		\caption{Beam, target, and detector main characteristics for the experiments considered in this work, along with the total number of protons on target (PoT) and the typical lower energy cut. The distance (to the centre of the experiment) $D$ and the typical detector dimensions (length $L$ and cross-area $S$) are also indicated. Note that the SHiP design is not final. We list the number of events (NoE in the table) corresponding to a $90 \%$ confidence level exclusion limit. In the DUNE-PRISM case, we considered both an on-axis and an off-axis configuration (see text for details). The references in the first column refer either to a published analysis in 
		the case of existing constraints, or to projected bounds in the case of planned experiments. }
		\label{tab:exp} 
\end{table*}

In this section, we present the revised exclusion limits and we discuss the estimates of the 
sensitivities that we have obtained for a representative selection of existing and planned proton beam-dump experiments, after the new positrons annihilation production mechanism is included 
in the evaluation of the LDM  yield. For each case we briefly discuss the relevant experimental details, and the assumptions  made in carrying out the analysis(see also Tab.~\ref{tab:exp}). Our results are  summarised in Sec.~\ref{sec:results}.

\subsection{MiniBooNE}

MiniBooNE is a proton beam-dump experiment at Fermilab, originally designed to measure short-baseline neutrino oscillations~\cite{Aguilar-Arevalo:2013pmq}. The MiniBooNE detector is a 6~m radius spherical tank, filled with 818 tons of mineral oil~\cite{AguilarArevalo:2008qa}. It is installed approximately 540~m downstream of a beryllium neutrino production target, where the 8~GeV proton beam from the Fermilab Booster impinges on. 

Recently, a dedicated LDM measurement was performed by the MiniBooNE-DM collaboration using data corresponding to $1.86\cdot10^{20}$ protons on target~\cite{Aguilar-Arevalo:2018wea}. Since neutrino interactions in the detector represent an irreducible background for the LDM measurement, the experiment was performed by steering the primary proton beam in an ``off-target'' configuration, to avoid 
neutrino production in the target, and to impinge directly on the steel beam-dump installed 50~m downstream. This resulted in a neutrino background reduction of a factor $\simeq 30$. The experiment considered both the nucleon and the electron scattering channel to detect LDM, with the latter providing the most stringent limits. 
After employing a sophisticated set of selection cuts to discriminate between the LDM signal and the residual backgrounds, zero events were observed in the signal region. This allowed the collaboration to set a 90$\%$ CL limit on the LDM parameters space, corresponding to 2.3 expected signal events. 

To compute the LDM flux in MiniBoone, we described the beam dump in \texttt{\geant} as a 4~m long steel block. Since this correspond to approximately 24 hadronic interaction lengths, we ignored any further downstream material. Also, we did not include any material upstream the thick target. In this work, we only considered the $\chi-e^-$ scattering process. We reproduced the MiniBooNE-DM analysis following the same strategy adopted in Ref.~\cite{deNiverville:2016rqh}. We parametrized the MiniBooNE-DM response with the following selection cuts, $E_e > 75$ MeV and $\cos(\theta_e)>0.99$, where $E_e$ and $\theta_e$ are, respectively, the scattered electron energy, and the angle measured with respect to the primary beam direction. The validity of this parametrisation can be assessed from Fig.~\ref{fig:ReachMB}, where we compare the sensitivity for the ``traditional'' LDM production as reported by the MiniBooNE-DM collaboration (dashed orange line)
with that obtained applying the aforementioned selection cuts (solid rust line) observing a very good agreement.

\subsection{NO$\nu$A}

NO$\nu$A is a neutrino experiment at Fermilab studying the oscillation of muon neutrinos to electron neutrinos~\cite{Acero:2019ksn}. The experiment measures neutrinos produced in the NuMI target facility by the 120~GeV proton beam from the FNAL Main Injector~\cite{Adamson:2015dkw}. The NO$\nu$A near detector (NO$\nu$A-ND) is located 990~m downstream from the target, at $14.6$~mrad angle from the primary beam direction. Such off-axis configuration was chosen to optimise the neutrino energy distribution for the oscillation measurement. The detector is a large volume of plastic (PVC) extrusions filled with liquid scintillator (active volume), followed by a muon detector made of alternating steel planes and scintillator planes. The active volume is a high-granularity sampling calorimeter, characterised by enhanced PID and tracking capabilities. The corresponding mass is approximately $193 \cdot 10^{3}$~kg, for a total volume of $3.9 \times 3.9 \times 12.67$~m$^3$ \cite{Acero:2019qcr}. 

A first estimate of the NO$\nu$A-ND sensitivity to LDM was discussed in~\cite{deNiverville:2018dbu} where, however, only the  $\chi-e^-$ scattering channel was considered. This result was based on a preliminary report of the $\nu-e$ elastic scattering analysis performed by the collaboration~\cite{Bian:2017axs}, for a total exposure of $2.97 \times 10^{20}$ POT. Both the elastic neutrino scattering signal (120 expected events) and the corresponding backgrounds (40 expected events) were treated as an irreducible background for the LDM search, for a 90$\%$ CL exclusion limit of $\simeq$ 16.4 LDM events.

In this work, we computed the NO$\nu$A-ND sensitivity to LDM by simulating electrons- and positrons-induced production processes in the NUMI target. We implemented the official \geant description of the target geometry and materials, 
as was used to measure fundamental neutrino properties~\cite{Acero:2019ksn}, and 
that was provided to us by the NO$\nu$A collaboration. 
We considered the NO$\nu$A-ND active volume described before, with an average electron number density $n_e\simeq 3\cdot 10^{23}$ cm$^{-3}$. Finally, we parameterized the detector response to the scattered electron with the following selection cuts: $E_e > 500$ MeV, $E_e \cdot \theta_e^2 < 5$ MeV, where $\theta_e$ is measured with respect to the impinging particle direction.

\subsection{SHiP}

SHiP is a proposed beam-dump experiment at CERN SPS to search for weakly interacting long lived particles~\cite{Anelli:2015pba}. The SHiP detector, currently being designed, foresees two complementary apparatus, to investigate the hidden sector exploiting both the visible decay signature of hidden particles and the recoil signal from the scattering on atomic electrons and nuclei. In particular, the SHiP Scattering and Neutrino Detector (SND) is a hybrid apparatus consisting of alternating layers of an absorber, nuclear emulsion films and fast electronic trackers, characterized by a very low detection threshold and enhanced PID capability. The detector is located approximately 40~m from the production target where the 400~GeV proton beam impinges on.

A first estimate of the SHiP experiment sensitivity to LDM was discussed in~\cite{deNiverville:2016rqh} considering both the $\chi-e^-$ and the $\chi-N$ scattering processes. More recently, the SHiP collaboration presented an updated limit for the $\chi-e^-$ channel, based on a robust evaluation of the irreducible neutrino background and on a realistic parameterization of the foreseen detector response, for a total exposure of $2\cdot 10^{20}$ POT~\cite{Ahdida:2654870}. 

In this work, we evaluated the SHiP sensitivity to LDM as follows. We computed the LDM flux due to positrons annihiliation in the beam dump with \texttt{\geantnospace}, implementing the current target geometry and material composition that were provided to us by the collaboration. We parameterized the SND active volume as a $90 \times 75 \times 320$ cm$^3$ volume, located 38~m from the beam dump, with a fiducial mass of 10~ton. The following selection cuts were applied to the scattered electron kinematics, 1~GeV $<E_e< $ 20~GeV, 10~mrad $<\theta_e<$ 20~mrad, with $\theta_e$ measured with respect to the impinging LDM particle direction. Within this signal region, we assumed an irreducible neutrino background of $800$ events~\cite{Ahdida:2654870}. This corresponds to a 90$\%$ CL exclusion limit of $\simeq 38$ events.

\subsection{DUNE}

DUNE is a large-scale experiment under construction in the US conceived for neutrino and proton decay studies~\cite{Abi:2020evt}. 
DUNE will consist of a  near detector, that will record interactions near the source of the beam, and of a much larger far detector, located underground 1,300~km downstream of the source. 
DUNE will detect neutrinos produced by the primary 120~GeV proton beam of 
the Fermilab accelerator complex impinging on a graphite target. 

In a recent work it was shown that, despite the abundant neutrino background, a dedicated analysis with the DUNE near detector data will be able to explore unknown territories in the LDM parameters space, exploiting the $\chi-e^-$ scattering channel~\cite{DeRomeri:2019kic}. 
In this work, we adopted the same description for the DUNE near detector geometry used in Ref.~\cite{DeRomeri:2019kic}, considering a 3x4x5 m$^3$ liquid argon detector located 574~m downstream from the target. 
We described the target as a thin, 220-cm long graphite cylinder~\cite{Papadimitriou:2018akk}. We parameterized the detector response with the following cuts on the scattered electron kinematics: $E_e \theta_e^2 < 2 m_e$, $E_e > 50$ MeV, with $\theta_e$ measured with respect to the impinging $\chi$ direction.

To derive the DUNE near detector exclusion limits for LDM, we considered a total accumulated charge of $1.1\cdot10^{21}$ POT/year, and a 7-years long measurement. We observe that, as discussed in Ref.~\cite{DeRomeri:2019kic}, the DUNE near detector sensitivity to LDM can be significantly enhanced by performing multiple measurements at different off-axis locations, to exploit the different angular spectra of the LDM signal and the neutrino background (DUNE-PRISM detector concept). In this work, for simplicity we performed a first estimate of the DUNE sensitivity to LDM produced by secondary $e^+$ considering both a single on-axis and a single off-axis measurement (at the maximum transverse distance of $36\,$m), leaving a more comprehensive evaluation for the future. We estimated the irreducible neutrino background for the on-axis (off-axis) measurement to be $\simeq 71\cdot10^{3}$ ($\simeq 1500$) events, assuming an equal experiment run time in neutrino and anti-neutrino mode~\cite{DeRomeri:2019kic}. This corresponds to a 90$\%$ CL exclusion limit of $350$ ($54$) signal events.

\subsection{Results}\label{sec:results}

In this section we present our results for the limits and for the projected sensitivities 
of the four experiments described above,  assuming that LDM is a complex scalar particle, 
that is for the model discussed in Sec.~\ref{sec:model}. In order to consistently compare the dark matter production via meson decay and via resonant production in the electromagnetic shower, we have used for the former the $\pi^0$ and $\eta$ meson yields from the \texttt{\geant} simulation described in the previous sections. Similarly the assumptions on the detectors geometry, signal response, and backgrounds have been applied to both type of production.

In the following, limits denoted as  $\varepsilon_{\rm lim}^{e}$ 
are based only on  $e^+/ e^-$ processes, that is they are derived considering 
only  dark photon  interactions with the leptons.  They can therefore also be used to constrain protophobic dark matter scenarios, for which proton beam-dump experiments are 
usually believed to have no sensitivity. For the coupling   $g_e$ of a dark photon interacting dominantly with the leptons and with suppressed couplings to hadrons, the limits on the couplings are given by the simple relation:
\begin{align}
 g_e^{\rm lim} = e\,  \varepsilon_{\rm lim}^{e} \ .
\end{align}
In the following figures, this ``lepton-only'' limit $\varepsilon_{\rm lim}^{e}$ is represented as a solid green line. Note that being electron-based experiments, the limits from NA64 and BaBar also apply in this case.

\begin{figure}
	\centering
		\includegraphics[width=0.48\textwidth]{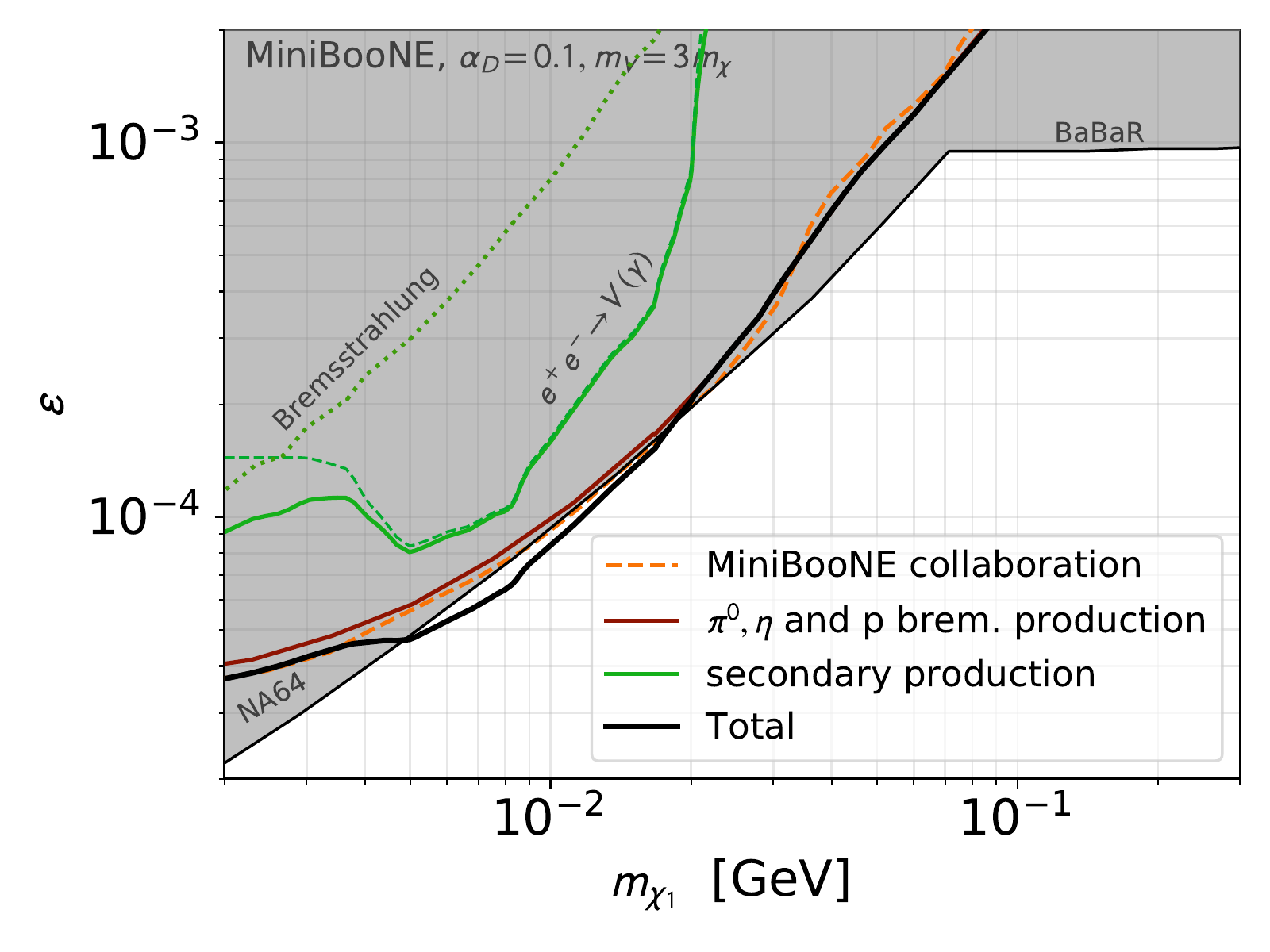}
	\caption{Limits for the MiniBooNE experiments.
	The grey region represents the exclusion bounds from the BaBaR~\cite{Lees:2017lec} and NA64~\cite{NA64:2019imj} collaborations. The dashed orange line corresponds to the sensitivity as extracted from~\cite{Aguilar-Arevalo:2018wea}, the rust solid line is our estimate based on hadronic processes only, the solid green line is our estimate based on secondary production processes only, and the thick black line is the combination of the two. 
	}
			\label{fig:ReachMB}
\end{figure}

We first considered the reach of the MiniBooNE experiment. As can be seen in Fig.~\ref{fig:ReachMB}, we find excellent agreement between our simulation using light meson production (orange dashed line) and the original limit  from the collaboration~\cite{Aguilar-Arevalo:2018wea} (rust solid line). This confirms the robustness of our calculations. The dotted and  dashed  green lines correspond, respectively, to the limits from bremsstrahlung, and from
positron-induced production, including both resonant $e^+ e^- \to V \to \chi \chi$ and associated $e^+ e^- \to \gamma V \to \gamma \chi \chi$ processes. They contribute significantly to the total number of expected events for $m_V \sim 20 $ MeV, thus significantly enhancing the full MiniBooNE exclusion limits compared with those from the NA64 collaboration. The mass range where the pure resonant process is active is clearly visible in the plot. In particular, the lower bound at $M_{\chi_1} \sim 3$ MeV ($m_V \sim 10$ MeV) is due to the fact that, following Eq.~\eqref{eq:mth}, a dark photon resonantly produced at this low mass does not transfer enough energy to the LDM particle (and ultimately to the scattered electron) to pass the  $E_{\rm th}$ selection cut. For dark photon masses below this threshold, the dominant production processes are thus the dark bremsstrahlung from electrons and positrons and the associated dark photon production. Note that the limit from secondary production is conservative in that we do not include dark photon production via the Compton-like process $\gamma e^- \to V e^-$.\footnote{As shown in Appendix~\ref{sec:app3}, based on the similarities with the  associated production differential cross section it is possible to estimate the typical size of the complete secondary production rate by multiplying by $\sim 3$ the associated production rate. Such modification, however, improves only marginally the limits presented here. We thus leave a complete study of the Compton-like process for a future work.}  For the lowest dark photon mass, as discussed in Appendix~\ref{sec:app2} and~\ref{sec:app3}, the cross-section for bremsstrahlung increases quadratically with the inverse of the dark photon mass, while associated production saturates.

\begin{figure}
	\centering
	\includegraphics[width=0.48\textwidth]{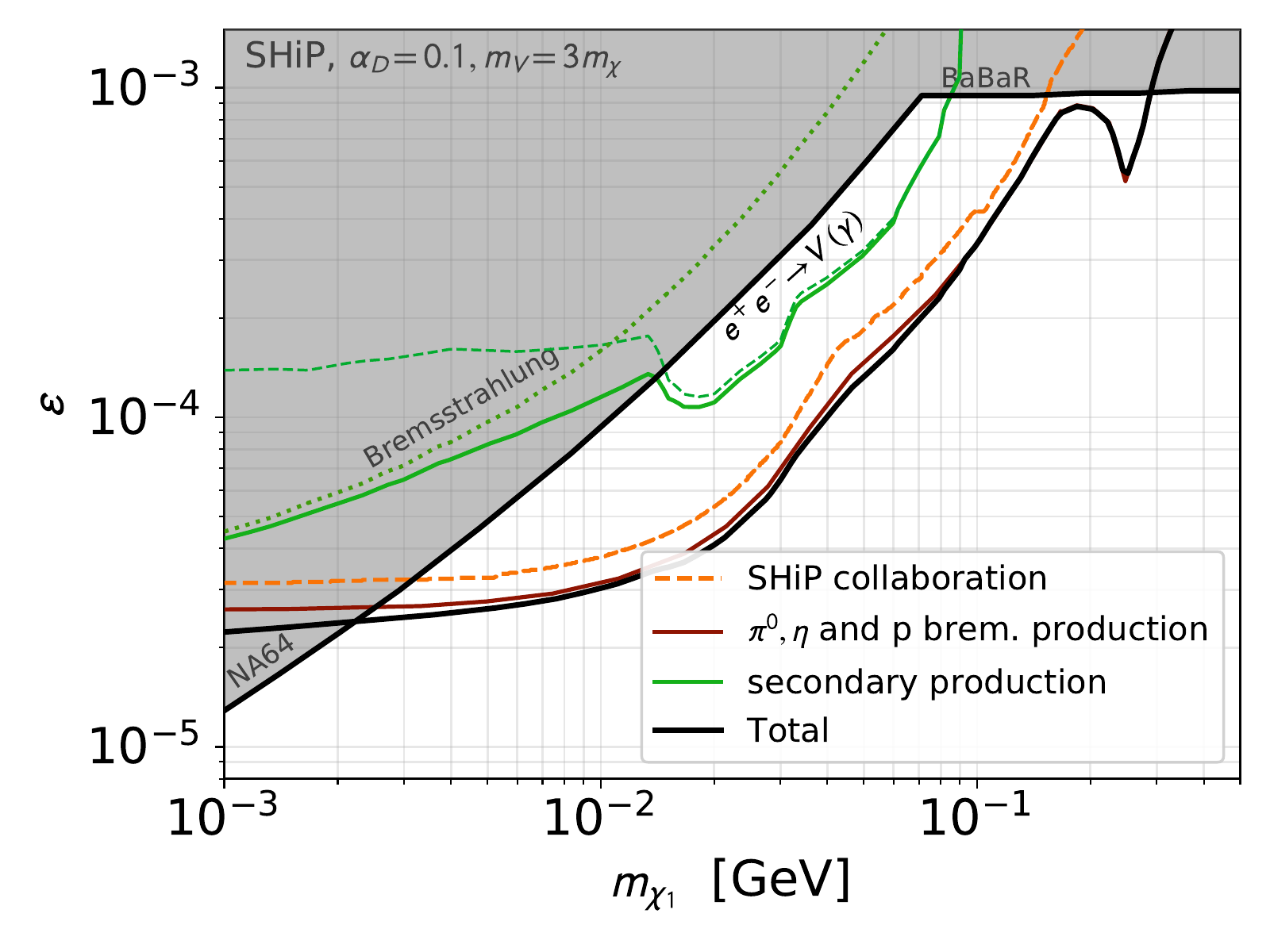}
	\caption{Projected reach of the SHiP experiment. The grey region represents the exclusion bounds from the BaBaR~\cite{Lees:2017lec} and NA64~\cite{NA64:2019imj} collaborations. The dashed orange line is the limit extracted from~\cite{SHIPtalk}, the rust line our estimate based on hadronic processes only, the solid green line our estimate based on the secondary production processes only, and the thick black line is the combination of both.}
	\label{fig:ReachSHIP}
\end{figure}

The impact of the energy threshold on the limit is further visible in Fig.~\ref{fig:ReachSHIP}, where we plot  the  expected sensitivity of the SHiP experiment. Also in this case, the comparison between our calculation (rust dashed line) and the results of the collaboration (orange dashed line) for light mesons LDM production show a relatively good agreement (notice that we did not include possible detection efficiencies in our estimate). Even if the experiment will use the $400$ GeV SPS proton beam, leading in principle to high-energy electromagnetic showers, due to the high detection threshold ($\sim 1$ GeV) electrons- and positrons-induced processes represent only a small fraction of the final events. 

\begin{figure}
	\centering
	\includegraphics[width=0.48\textwidth]{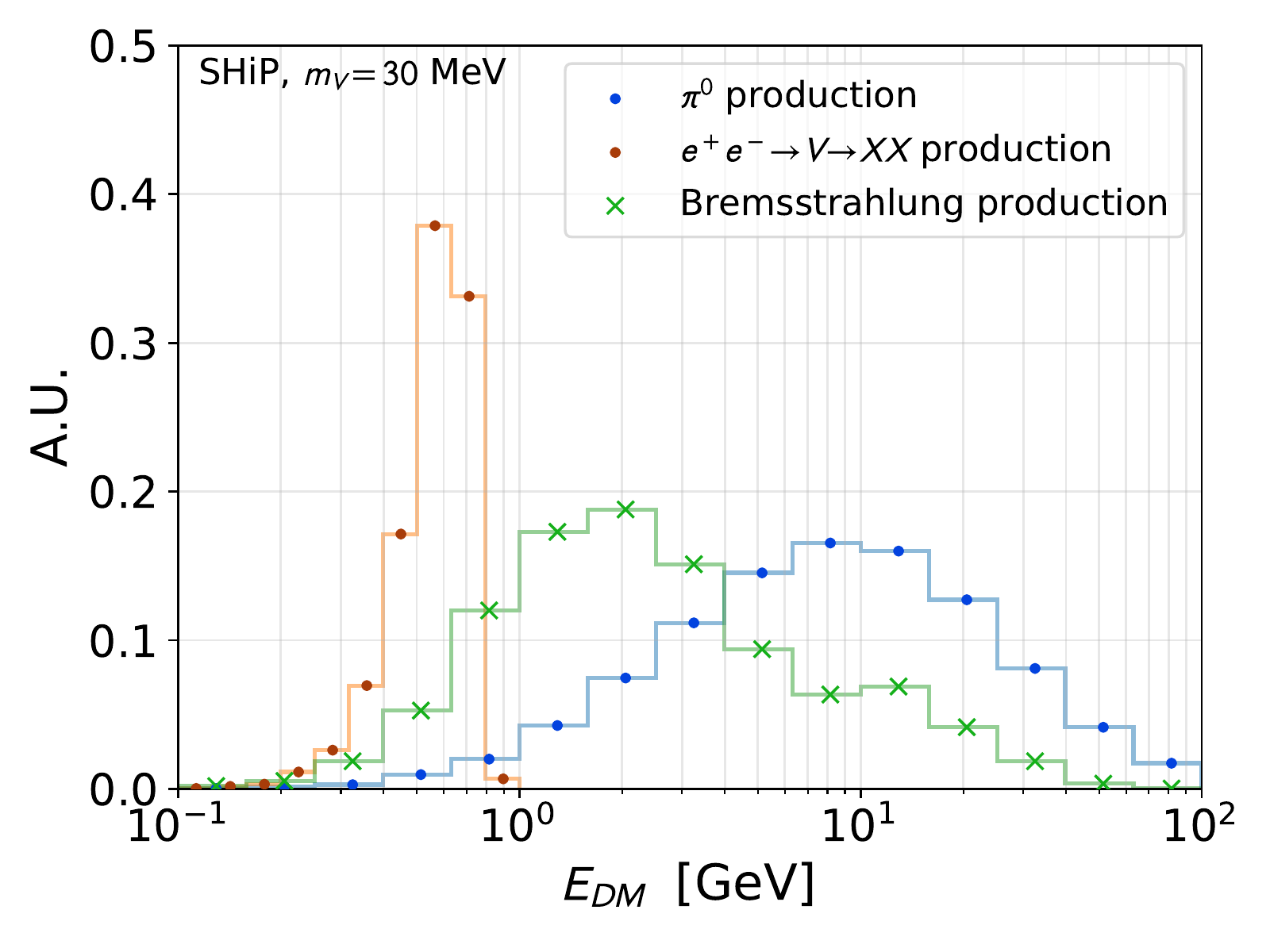}
	\caption{Energy distribution of LDM particles impinging on the SHiP detector for different production mechanisms: production from mesons decay (blue), positrons resonant annihilation (orange), electrons and positrons bremmstrahlung (green).}
	\label{fig:ESHIP}
\end{figure}

We show in more detail in Fig.~\ref{fig:ESHIP} the LDM energy distribution for the different production mechanisms, for the specific choice $m_V=30$ MeV. The energy distribution for the leading mesons decay channel peaks as the highest energies, as expected since it originates from mesons from the primary hadronic shower. The secondary production from electrons/positrons bremsstrahlung retains a significant fraction of the energy of the shower and peaks just above the GeV. As shown in Appendix~\ref{sec:app2}, this is due to both the fact that bremsstrahlung dark photons typically retain all the energy of the incoming $e^+/e^-$ and that the bremsstrahlung process itself is effective at large center-of-mass energy. Finally, LDM production through resonant positrons annihilation is peaked at a lower energy below the GeV, around half the energy of the outgoing dark photon $E_V = m_V^2/(2 m_e) \sim 0.9$ GeV.

\begin{figure}
	\centering
		\includegraphics[width=0.48\textwidth]{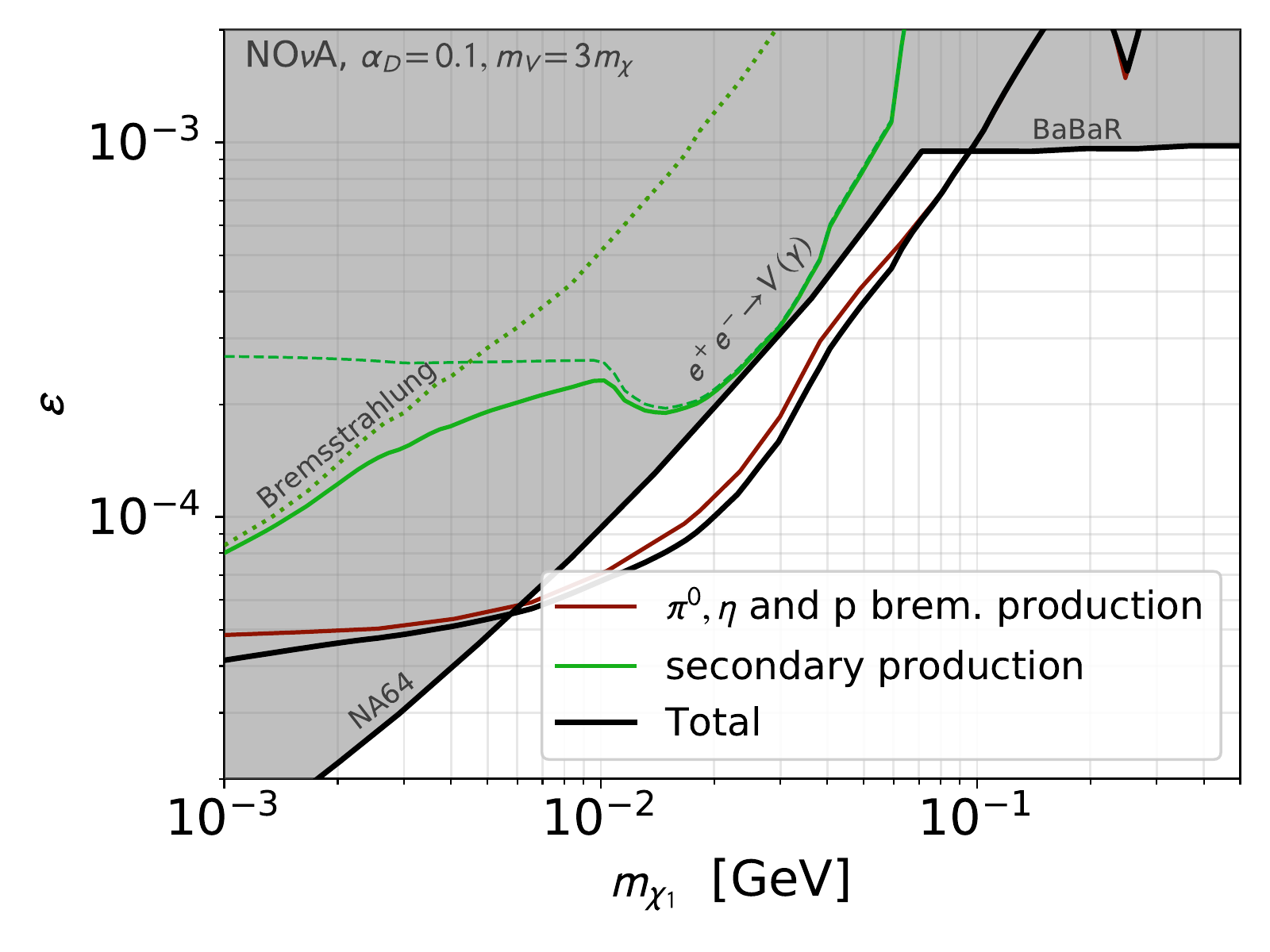}
	\caption{Projected reach of the NO$\nu$A experiment. The grey region represents the exclusion bounds from the BaBaR~\cite{Lees:2017lec} and NA64~\cite{NA64:2019imj} collaborations. The rust line is our estimate based on hadronic processes only, the solid green line our estimate based on secondary production processes only, and the thick black line is the combination of both.}
			\label{fig:ReachNOVA}
\end{figure}

\begin{figure}
	\centering
		\includegraphics[width=0.48\textwidth]{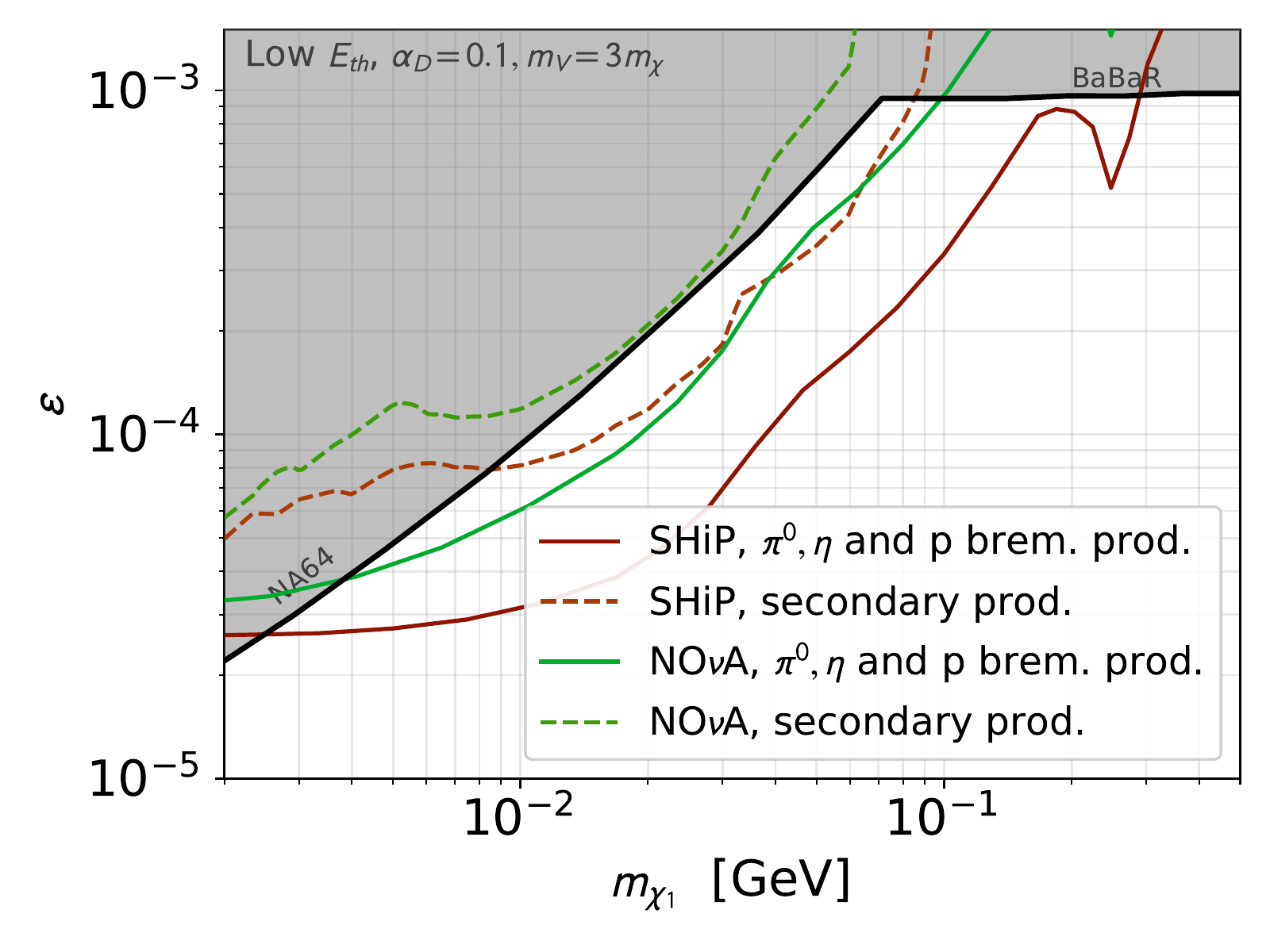}
	\caption{Projected reach of the NO$\nu$A (green lines) and the SHiP (red lines) experiments, with reduced energy thresholds at $125$ MeV for NO$\nu$A and $250$ MeV for SHiP. Sensitivity estimates are based on on $16.4$ ($38$) signal events for NO$\nu$A (SHiP). The grey region represents the exclusion bounds from the BaBaR~\cite{Lees:2017lec} and NA64~\cite{NA64:2019imj} collaborations. The solid lines are our estimate based on hadronic processes only, while the dashed lines are based on secondary production processes.}
			\label{fig:ReachLowEth}
\end{figure}

In the case of the NO$\nu$A experiment, the large energy threshold $E_{\rm th} = 0.5$ GeV also limits significantly the contribution of electromagnetic shower-induced processes, with a corresponding lower mass threshold around $M_{\chi_1} \sim 10$ MeV ($m_V \sim 30$ MeV), as seen in Fig.~\ref{fig:ReachNOVA}. Note that the relatively large energy threshold as well as the large distance between the beam dump and the experiment tends to reduce the contribution from the shower-generated events, since they are typically both less collimated and less energetic than their hadronic-generated counterparts.
We illustrate the effect of lowering the energy threshold for the NO$\nu$A and SHiP experiments in Fig.~\ref{fig:ReachLowEth}.  In this case, we did not combine the hadronic and leptonic limits as for the other plots, to illustrate that the background level are likely to be significantly modified, so that the proposed reaches should also be rescaled accordingly. On the other hand, it is clear that the ratios between both production modes is not significantly modified by this change. In particular, in the case of the NO$\nu$A experiment, the small geometric acceptance of the experiment suppresses naturally the shower-induced events. 

\begin{figure}
	\centering
		\includegraphics[width=0.48\textwidth]{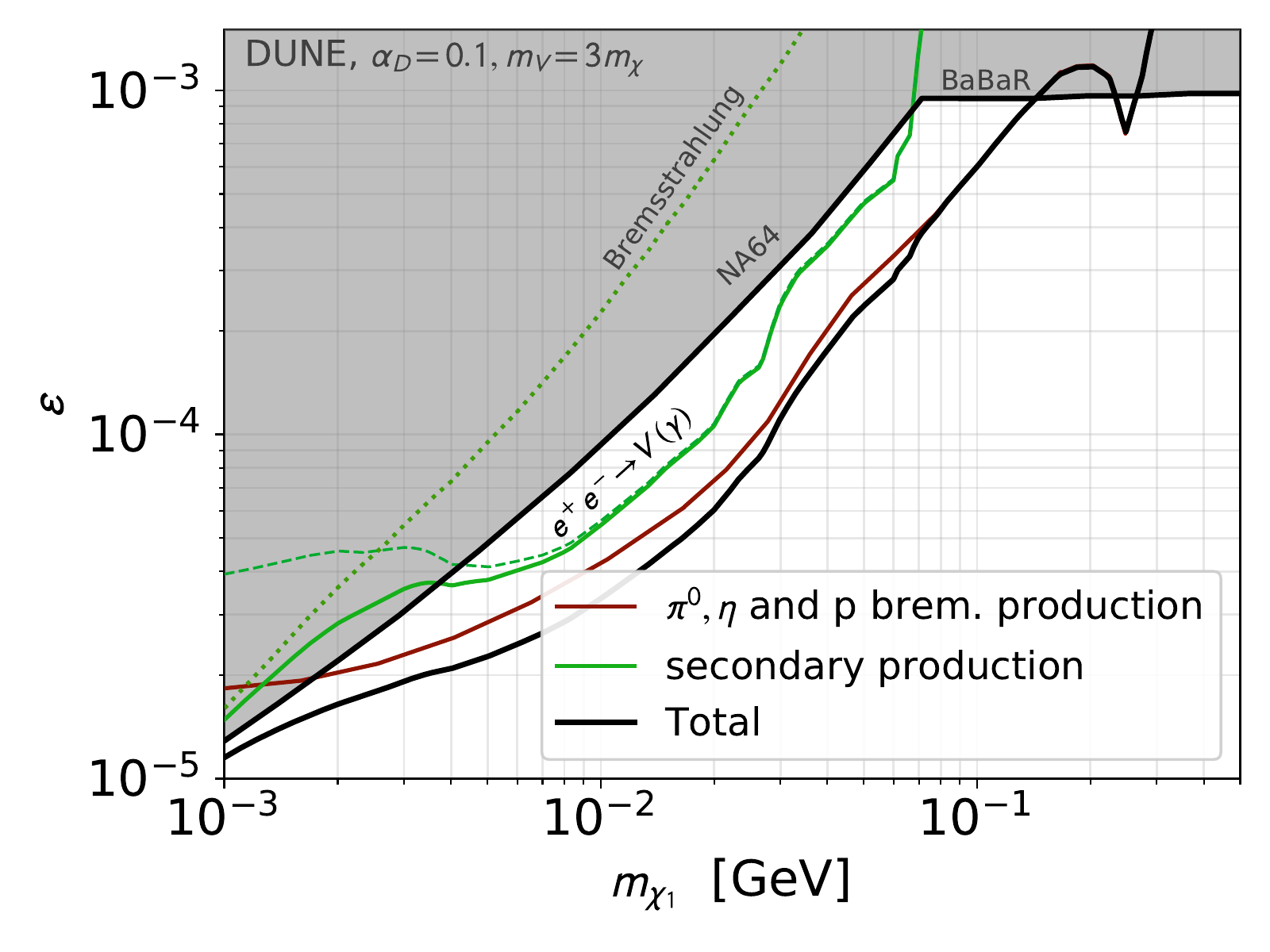}
	\caption{Projected reach of DUNE near detector. The grey region represents the exclusion bounds from the BaBaR~\cite{Lees:2017lec} and NA64~\cite{NA64:2019imj} collaborations. The rust line represents our estimate based on hadronic processes only, the solid green line our estimate based on secondary production only, and the thick black line is the combination of both.}
			\label{fig:ReachDUNE}
\end{figure}

\begin{figure}
	\centering
		\includegraphics[width=0.48\textwidth]{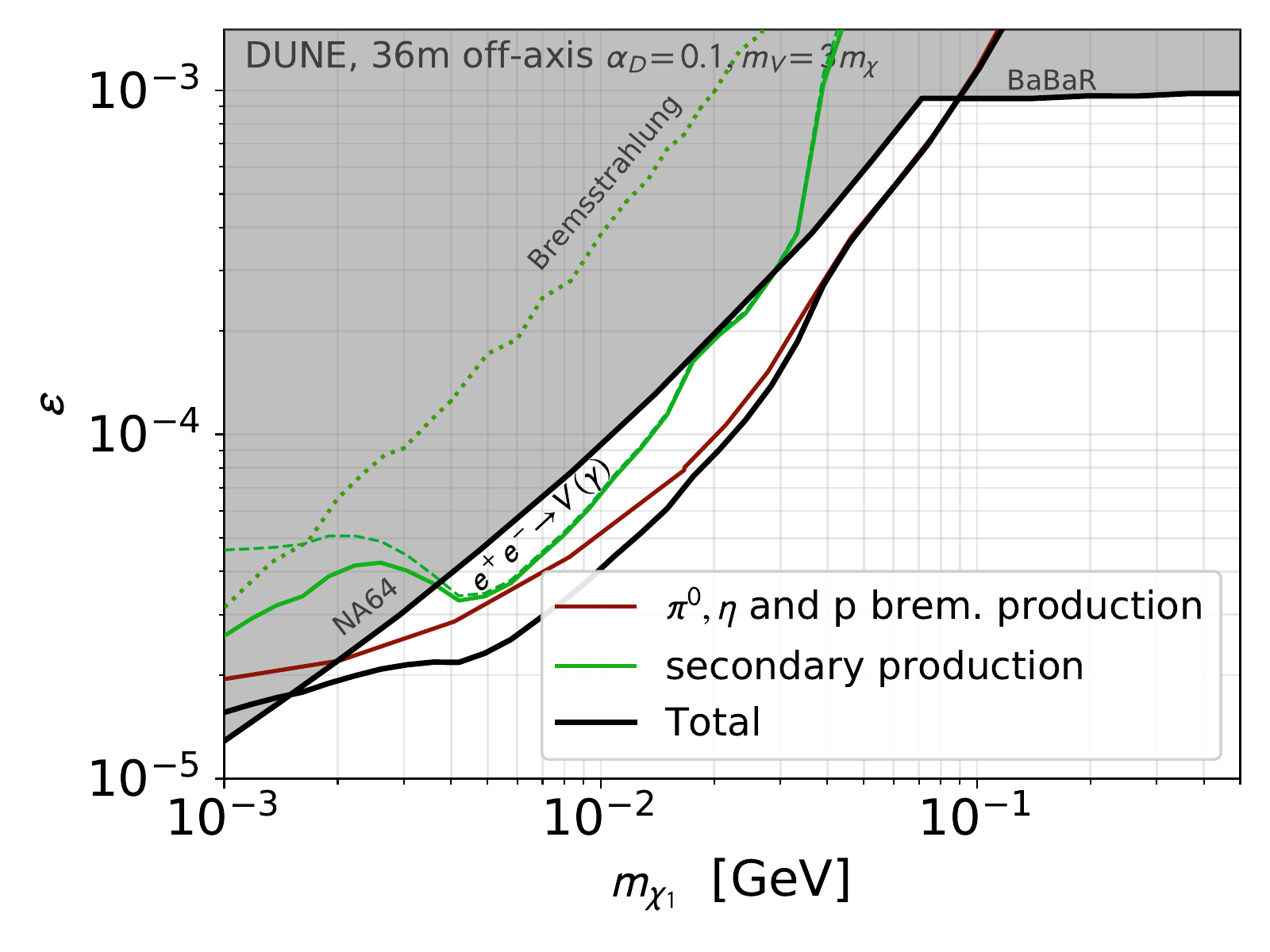}
	\caption{Projected reach of DUNE near detector, if moved by 36 m off the beam axis. Same color coding as the previous DUNE plot.
	}
			\label{fig:ReachDUNEOA}
\end{figure}

Finally, we present in Fig.~\ref{fig:ReachDUNE} the long term prospect based on the near detector of the DUNE experiment. This experiment will adopt a much lower energy threshold than NO$\nu$A and SHiP. 
Consequently, we observe that the leptonic-induced events play an important role in the final production rates, particularly at small dark matter masses.
A particularity of the proposed DUNE-PRISM near detector concept is that it can be physically moved off-axis up to $36$ m to reduce the overall background. While we did not performed a complete analysis like the one carried out in Ref.~\cite{DeRomeri:2019kic}, we present in Fig.~\ref{fig:ReachDUNEOA} the possible reach of the DUNE near detector in case it will be moved at the maximal off-axis distance, considering the same run parameters as the nominal on-axis mode. Interestingly, the wide emission cone of the leptons-induced dark matter candidate enhances their importance with respect to the standard mesons decay processes.

\section{Conclusions and outlooks}

When a high-energy proton beam impinges on a thick target, a large fraction of the primary energy is transferred to the electromagnetic component of the developed particles shower, resulting into an abundant production of photons, electrons, and positrons. In this work, starting from this observation, we have discussed for the first time the role of electrons- and positrons-induced processes in proton beam-dump experiments in relation to LDM searches. We have shown 
that LDM production from shower induced electromagnetic processes, that was so far 
overlooked, must be accounted for to properly assess the sensitivity of 
forthcoming proton-beam dump experiments, and to derive limits on the LDM parameter 
space from the analysis of existing data.

A numerical procedure, based on the \texttt{MADDUMP} and \texttt{BdNMC} simulations codes was developed to generate LDM particles, 
and, starting from the $e^+/e^-$ differential track length in the target computed with a \texttt{\geantnospace}-based simulation, to propagate them into a downstream detector. 
We considered a representative set of proton thick-target experiments (MiniBooNE, NO$\nu$A, SHiP, and DUNE), finding that for each of them the new production mechanism results into a non-negligible increment of the sensitivity to LDM. For some regions of the parameters space, the $e^+/e^-$-induced processes actually represent the dominant production mechanism for LDM, and can lead to signal rates on par with the standard results. Due to the typically softer spectrum of LDM particles generated from $e^+/e^-$ secondaries with respect to those originating from mesons decays, this effect is more important for experiments characterised  by low detection threshold on the scattered electron. 

Before concluding, it should be emphasised that, while we focused on the case of a dark photon mediator, our analysis can be easily extended to any other LDM model. Given that the increase in the LDM particle yield that we obtain only depends on the inclusion of new production channels, our results can be relevant also for LDM searches based on detection strategies different from the simple $\chi e^- \rightarrow \chi e^-$ scattering considered here, as for example measurements of energy deposition in the detector from visible decays of long-lived dark sector states.
Finally, while we concentrated on proton beam-dump experiments, it would  also be important to  properly account for the new processes analysed in this work for projected LHC-based intensity frontier experiments, such as FASER($\nu$)~\cite{Feng:2017uoz,Abreu:2020ddv}, MATHUSLA~\cite{Curtin:2018mvb}, Codex-b~\cite{Aielli:2019ivi}, ANUBIS~\cite{Bauer:2018onh} or MilliQan~\cite{Ball:2016zrp}. The extremely high energy available at LHC interaction points may actually lead to an even stronger production of dark sector particles from processes induced by electromagnetic showers. We thus believe that it would be  particularly important for  these experiments to  consider carefully also  shower-based dark sector productions, and not only to estimate correctly their sensitivity reach, but also to optimise the choice of the detection energy thresholds for the physics run.

\bigskip
\noindent \textit{Note added:} Simultaneously with our paper, Ref.~\cite{Dutta:2020vop} appeared which dealt with neutrino experiments based on high-intensity proton beam with$~\sim$ GeV energy (such as the COHERENT experiment) and investigated in details the use of timing and energy cuts to reduce the neutrino background. We point out that it would be interesting to include the complete shower productions modes (in particular resonant production) when estimating the efficiency of this approach. Indeed, the kinematic distribution of these events is likely to be significantly different from the meson-induced production, potentially leading to new ways of optimising the selection cuts.
\medskip

\bigskip
\medskip
\noindent \textbf{Acknowledgments}
\medskip

AC and LD warmly thanks L. Buonocore for very helpful discussions on \texttt{MADDUMP}. AC thanks P. Snopok and A. Habig for their support in implementing the official NUMI target geometry in our Monte Carlo simulation, W. Bonivento and T. Ruf for their support in implementing the official SHIP target geometry in our Monte Carlo simulation, and K. Kelly for his help concerning the neutrino background expected in the DUNE-PRISM measurement. The authors thank S. Trojanowski for his help on analtyical expression for the bremsstrahlung process.
LD and EN are supported by the INFN “Iniziativa Specifica” Theoretical Astroparticle Physics (TAsP-LNF).

\newpage

\appendix

\addcontentsline{toc}{section}{Appendices}

\section{Analytical treatment of EM showers}
\label{sec:app1}
In this Appendix we describe the technical details of the analytical shower modelling.
Our treatment is based on the study of the development  of high
energy cosmic ray showers in the atmosphere presented in Ref.~\cite{Lipari:2008td},
which in turn is based on the  Rossi and Griesen approach~\cite{RevModPhys.13.240}.

The idea is to solve first the equations coupling the differential density of electrons/positrons $\ne (E,t)$ and of photons $\ng (E,t)$ as function of the depth parameter $t$ (expressed in unit of radiation length), that read:
\begin{align}
\frac{\partial \nez (E)}{\partial t} &=
-\int_0^1 dx \left[ \frac{d\sigma^{b}}{dx} \left( \nez(E)- \frac{\nez(E/(1-x))}{1-x} \right) \right.  \nn 
\\
& \left. \qquad \qquad \qquad -  \frac{2}{x} \frac{d\sigma^{p}}{dx} \ng(E/x)  \right] 
\label{eq:showereq}
\\
\label{eq:showergamma}
\frac{\partial \ng (E)}{\partial t} &= -\sigma^{p} \ng + \int_0^1 dx \left( \frac{\nez(E/x)}{x} \frac{d\sigma^{p}}{dx}  \right)\,, 
\end{align} 
where $ \frac{d\sigma^{b}}{dx}$ and $\frac{d\sigma^{p}}{dx}$ are respectively 
 the differential cross section for 
bremsstrahlung photon production and for $e^\pm$ pair production, 
and  $\sigma^{p} = \int_0^1 dx  \frac{d\sigma^{p}}{dx}$  is the integrated pair production cross section. The two differential cross sections are given by:
\begin{align}
 \label{eq:dsigmab}
 \frac{d\sigma^{b}}{dx} (x) &= \frac{1}{x} \left[ 1 - \left( \frac{2}{3} -2 b_Z \right) (1-x) + (1-x)^2 \right] \\
 \label{eq:dsigmap} 
  \frac{d\sigma^{p}}{dx} (x) &= (1-x)^2 + \left( \frac{2}{3} -2 b_Z \right) (1-x)x +x^2 \ .
\end{align}
The first two terms in Eq.~\eqref{eq:showereq} represent respectively the fraction of $e^\pm$ of energy $E$ which loose energy by bremsstrahlung, and the fraction of higher energy $e^\pm$ which end up with energy $E$ following a bremsstrahlung. The last term accounts for
$e^\pm$ produced via photon conversion. 
The two terms in Eq.~\eqref{eq:showergamma} represent, respectively, the photons lost to pair-production and the photons produced via bremsstrahlung. The parameter $x$ represents the energy ratio $E_e/E_\gamma$ between the incident $e^\pm$ and the outgoing photon for  bremsstrahlung, 
while it represents the opposite ratio for $e^\pm$ pair production. The effective parameter $b_Z$ can be expressed as function of the atomic number $Z$ of the medium  as 
\begin{align}
    b_Z \simeq \frac{1}{18 \log (183 ~Z^{-1/3})} \ . 
\end{align}
As was worked out long ago by Rossi and Greisen~\cite{RevModPhys.13.240}, it is possible to obtain an analytical solution for the above set of coupled equations valid for the later stage of shower development, i.e. when $t, E/\Ez \ll 1$. For a shower induced by a photon of energy $\Ez$ the solution reads:
\begin{align}
\frac{d\nez (E, \Ez,t)}{dEdt} = \frac{1}{\Ez \sqrt{2 \pi} } \left[ \frac{G_{\gamma \to e} (s) }{\sqrt{\lambda_1''(s) t }} \left( \frac{E}{E_0 }\right)^{-1-s} e^{\lambda_1(s) t}\right] \ ,
\end{align} 
where we have used the primed notation for the derivatives with respect to $s$. The auxiliary function $G_{\gamma \to e} (s) $ is  defined as:
\begin{align}
 G_{\gamma \to e} (s) &=    - \frac{1}{C} \frac{[\sigma^p+\lambda_1(s)][\sigma^p+\lambda_2(s)]}{\lambda_1(s) - \lambda_2(s)} \,, 
 \end{align}
 while the two functions   $\lambda_{1,2} $ read:
 \begin{align}
 \lambda_{1,2}(s) &= - \frac{1}{2}(A + \sigma^p) \pm \frac{1}{2} \sqrt{ (A-\sigma^p)^2 + 4 B C } \ .
\end{align}
We have used the following cross-sections momenta:
\begin{align}
A(s) &= \int_0^1 dx \frac{d\sigma^b}{dx}  \left(  1 - (1-x)^s \right)\,,  \\
B(s) &= 2 \int_0^1 dx \frac{d\sigma^p}{dx}  x^s\,, \\
C(s) &= \int_0^1 dx \frac{d\sigma^b}{dx}  x^s \,, 
\end{align}
which can also be straightforwardly expressed as (lengthy) expressions involving polylogarithm functions~\cite{Lipari:2008td}. 

Once this un-cut distribution is estimated, the approach of Rossi and Griesen  is to add a ``loss'' term in Eq.~\eqref{eq:showereq} by replacing 
\begin{align}
    \frac{\partial \ne (E)}{\partial t} ~\to \frac{\partial \ne (E)}{\partial t} - \epsilon_c \frac{\partial \ne (E)}{\partial E}\,, 
\end{align}
where $\epsilon_c$ is the critical energy defined in Eq.~\eqref{eq:criticalE}. Approximate solutions to the new  system of equations can be searched for in the form:
\begin{align}
    \ne (E,s)  = \nez (E,s) \times p_1(E/\epsilon_c,s) \ .
\end{align}
In general $\nez (E,s) $ on the right-hand-side of this equation 
should be multiplied by a cut-off function $p$ that can in principle be obtained by replacing 
$n^0_e\times p$ in the system of 
differential equations.  In our paper, we are using for simplicity 
the interpolation of $p$, estimated at the shower maximum, that is 
$p\to p_1(x=E/\epsilon_c,s=1)$ as given in~\cite{Lipari:2008td}. Note that a good analytical interpolation in $x = E/\epsilon_c $ is given by:
\begin{align}
   p_1(x,1) = \tanh (1.8 x^{0.18})^{18}  \ .
\end{align}

Finally, since the original hadronic shower produces a large number of photons with different energy, the resulting track-length distribution for the full electromagnetic shower is obtained by integrating over the initial differential distribution of photons, as shown in Eq.~\eqref{eq:tlanalytical}.

\section{Numerical approach to bremsstrahlung processes}
\label{sec:app2}

Bremsstrahlung production of dark photons is traditionally the dominant production mechanism considered in electron beam-dump experiments. We give in this Appendix a few details about our estimation of this process via \amc, starting from a brief summary of the analytical approach based on the Weizsacker-Williams approximation~\cite{Tsai:1986tx,Bjorken:2009mm,Andreas:2012mt}.  
We present the result for the case of an incoming electron, but note that it also  
applies for the case of an incoming positron.

We consider the process
\begin{align}
e^- (p) N(P_i) \to e^-(p')  N (P_f) V^{(*)} (k)  \to e^- N \chi^* \chi \ ,
\end{align}
where $N$ is a nucleus with atomic number $Z$. For simplicity, we focus on the case of a monochromatic impinging beam (the extension to the realistic case through a track length approach is straightforward). We follow the notations and summarising the discussion of~\cite{Bjorken:2009mm}. 
We define as $E_0 (E_V)$ the energy of the incoming electron (outgoing dark photon) in the lab frame, and we introduce the ratio $x \equiv E_{V}/E_0$. As was noted in~\cite{Tsai:1986tx}, the photons mediating the process are only very mildly virtual so that their interaction with the electron are dominated by their transverse polarisation.  It is then possible to decompose the cross section into a real photon-electron
scattering, $e(p) \gamma(q) \rightarrow e(p') V(k)$ where the photon has the (small) virtual momentum $q \equiv P_i -
P_f$, and a form factor  for the emission of the photon from the nucleus. 
Let us define  $t \equiv -q^2$ (not to be confused with the the  depth  parameter $t$ introduced in the previous Appendix) and  
call $\theta_{V}$ the angle of the outgoing dark photon with respect to the incoming electron in the lab frame. The full cross section can be written~\cite{Tsai:1986tx}:
\begin{widetext}
\begin{equation}
\frac{d\sigma( 2 \to 3)}{d x 
  d\cos \theta_{V}}  = E_0  \left(\frac{\alpha_{\rm em} \mathcal{F}}{\pi}\right) \left(\frac{E_0 x
  \beta_{V}}{(1-x)}\right) \times \frac{d\sigma(p + q \rightarrow p' + k)}{d(p \cdot k)}\bigg|_{t=t_{\rm min}} ,
\end{equation}
\end{widetext}
with $\beta_{V} \equiv \sqrt{1-m_V^2/E_0^2}$.
Importantly, the cross section for the $2\to2$ process is estimated at the minimum virtuality $t=t_{\rm min}$. The term $\frac{\alpha_{\rm em} \mathcal{F}}{\pi}$ describes the effective photon flux integrated from $t=t_{\rm min}$ to the total center of mass (CM)
energy $t_{\rm max} = s$.  It can be obtained by integrating the nuclear and atomic form factors over the virtuality:
\begin{equation}
\mathcal{F} \equiv \int_{t_{\rm min}}^{t_{\rm max}}  dt \frac{t-t_{\rm min}}{t^2} G_2(t)\,,
\end{equation}
with $G_2(t) = G_2^{el} + G_2^{in} $ defined by
\begin{align}\nonumber
	G_2^{el}&=\left(\frac{a^2 t}{1 + a^2 t}\right)^2 \left(\frac{1}{1 +
	t/d}\right)^2 Z^2\,, \\
	G_2^{in}&=\left(\frac{{a^\prime}^2 t}{1+{a^\prime}^2 t}\right)^2 \left(\frac{1 +
		\frac{t}{4m_p^2}(\mu_p^2 - 1)}{\left(1 +
		\frac{t}{0.71\,\mathrm{GeV}^2}\right)^4}\right) Z \, ,
\end{align}
with $\mu_p = 2.79$ and the proton mass $m_p = 0.938$ GeV.\footnote{Note that the last term of the inelastic form factor is not squared, following the original expression of~\cite{Tsai:1986tx} (see also~\cite{Jodlowski:2019ycu}) compared to the expression in~\cite{Bjorken:2009mm}.} Interestingly, we see that the form factors disfavour very soft or very hard photon exchanges due to either the screening from the electrons in the atomic cloud when 
\begin{align}
a^2 t, a'^2 t \ll 1 , \ \  a \equiv 111 \frac{1}{m_e Z^{1/3}} ,\ \  a' \equiv 773 \frac{1}{m_e Z^{2/3}} 
\end{align} 
or from the finite nuclear size in the other limit
\begin{align}
d t \ll 1 , \qquad d = 0.164 ~\textrm{ GeV}^2 A^{-2/3} \ .
\end{align} 
As pointed out by~\cite{Bjorken:2009mm}, all values of $t$ contribute equally to the integral -- in particular, the integral it is not dominated by $t \sim t_{\min}$. Indeed, while the virtual photon propagator squared, $1/t^2$, is maximum at $t=t_{\rm min}$, the phase-space numerator balances it in the integral. The minimum value of $t$ is given by
\begin{align}
 t_{\rm min} = -q_{\rm min}^2 \approx \left(\frac{U}{2  (1-x)}\right)^2 \sim \left( \frac{M_V^2}{2 E_0}\right)^2,
\end{align}
where
\begin{align}
U & \equiv & U(x,\theta_{V}) = E_0^2 \theta_{V}^2 x + m_{V}^2 \frac{1-x}{x} + m_e^2 x \ .
\end{align}at $t \sim t_{\rm min}$ 
Following~\cite{Bjorken:2009mm}, the cross section for the $2 \to 2$ process 
at $t \sim t_{\rm min}$ 
can be written up to terms in $m_e^2$  as:
\begin{align}
\frac{d\sigma}{d(p \cdot k)} = 2 \frac{d\sigma}{d t_2} &=  (4\pi\alpha_{\rm em}^2\epsilon^2) \frac{(1-x)}{U^2} \left[ 1+(1-x)^2 \right. \nn \\
& \left. +\frac{2 (1-x)^2 m_{V}^2}{U^2} \Big( m_{V}^2 - \frac{U x}{1-x}\Big) \right].
\end{align}
Putting everything together and neglecting the $\theta_{V} $ dependence in $\mathcal{F}$, the cross section can be integrated once yielding
\begin{align}
\frac{d\sigma_{3\rightarrow 2}}{d x} = 4 \alpha_{\rm em}^3\epsilon^2 \mathcal{F} \beta_{V} \left( m_{V}^2 \frac{1-x}{x} + m_e^2 x \right)^{-1} \! \! \! \Big(1-x+\frac{x^2}{3}\Big)\,,
\end{align}
(note that the original expression from~\cite{Bjorken:2009mm} missed a factor of $1/2$ \cite{Andreas:2012mt}). It is clear that this differential cross section has an approximate singularity for $x \sim 1$, regulated by the electron mass at $(1-x)_{c_1} = \frac{m_e^2}{m_{V}^2}$, where the subscript $c_1$ labels a first cutoff point.  As remarked in~\cite{Bjorken:2009mm}, the approximation also breaks down if the virtuality is too large, yielding a second cutoff $(1-x)_{c_2} =
\frac{m_{V}^2}{E_0^2}$.   The total cross section finally reads:
\begin{align}
\sigma \approx \frac{4}{3} \frac{\alpha_{\rm em}^3\epsilon^2 \mathcal{F} \beta_{V}}{m_{V}^2} \, \log\left(\frac{1}{(1-x)_c }\right)\,, 
\end{align}
where $ (1-x)_c = \max\bigg( \frac{m_e^2}{m_{V}^2},\, \frac{m_{V}^2}{E_0^2}\bigg) $.

An important feature that can  be read out from this formula is that the cross section is actually only mildly dependent on the incoming electron energy, either via the logarithm term (which saturates when the $\frac {m_e}{m_{V}}$ contribution dominates), or via the form-factor contribution, which also saturates at high energy due to the atomic electrons screening. 

We have simulated this process in \amc using an effective $N N \gamma$ interaction with form factor $G_2$. This implies that we did not use the Weizsacker-Williams approximation for the cross section, but we directly estimated the $2 \to 4$ process with dark matter final states. Furthermore, in order to regulate the numerical divergence which arises for large electron energies when the exchanged photon is very soft, we have modified the form factor $G_2(t)$.
In particular, due to the screening effects occurring when $a^2 t \ll 1$, we know that this part of the phase space is sub-dominant in the final production rate. We therefore implemented a regularisation cut by setting the form factor to $0$ in the ``screened'' region:
\begin{align}
G_2^r (t) = \begin{cases} G_2 (t) &\text{   for } a^2 t > 1/3 \\ 0 &\text{   for } a^2 t < 1/3\ .
 \end{cases}
\end{align}
We have explicitly checked that the value of the final cross section is not modified by varying the cut between $a^2 t < 1$ and $a^2 t  < 0.05$, and agrees with the analytical expression developed above. Furthermore, we have verified that the differential distribution in angles and energy are also not affected by this regularisation procedure.

\section{Associated and Compton-like process}
\label{sec:app3}

We give in this Appendix more details about the associated production and Compton-like scattering which complement the pure resonant production of light dark matter. 

The differential cross section for both processes peaks forward at $\theta \sim 0$, with $\theta$ the $V$ production angle in the CM frame (although the associated production process is also enhanced in the opposite direction, $\theta \sim \pi$). For small angles and in the limit $\sqrt{s} \gg m_V, m_e$, the following similar expressions hold:
\begin{align}
\label{eq:dCSassoc}
\frac{d \sigma_{\rm assoc}}{d \cos \theta} = \frac{d \sigma_{\rm Compton}}{d \cos \theta} \simeq \frac{4 \pi \varepsilon^2 \aem^2 }{s \theta^2 + 4 m_e^2} \ ,
\end{align}
In particular, both differential cross sections saturate at very small angle, when $s \theta^2 < 4 m_e^2$. The total cross sections are also equivalent, with
\begin{align}
\sigma_{\rm assoc} \simeq \frac{2 \pi \varepsilon^2 \aem^2 }{m_e E_+} \left(\log (\frac{2 E_+}{m_e}) - 1\right) \\
 \sigma_{\rm Compton} \simeq \frac{\pi \varepsilon^2 \aem^2 }{m_e E_+} \left( \log (\frac{2 E_+}{m_e}) + \frac{1}{2} \right)\ ,
\end{align}
where the factor of $2$ is compensated by the fact that the associated production also generate efficiently events with a very forward photon, with  the same rate as in the forward dark photon region. Hence both processes lead to similar production rates of energetic dark photons, and since the cross section does not depend on $m_V$, we expect these rates to saturate in the light dark photon limit. Finally, note that we have considered for both processes the atomic electrons to be free (i.e. described by a plane wave wavefunction) and in particular we neglected the target electron motion~\cite{Nardi:2018cxi}.

Furthermore, we observe that in an electromagnetic shower, the distribution of photons actually follows relatively closely the one of the positron/electron as long as the energy is above the critical energy. One has $ T_\gamma \sim (1.3-1.5) \cdot (T_{e^+}+T_{e^-}) $ in most of the shower development -- see for example the discussion in Ref.~\cite{Lipari:2008td}. All in all, we therefore expect the production of very forward dark photons in the electromagnetic sub-shower to be a factor of $2$ larger for the Compton-like production than for the associated production, albeit with very similar kinematics. 

We have simulated the associated production process in \amc using the positron track length estimated via \texttt{\geant}. As can be seen from the differential cross section Eq.~\eqref{eq:dCSassoc}, the process has an approximate  collinear divergence regulated by the electron mass which leads to a logarithmic enhancement of the total cross section. 
We numerically-regulated this divergence in \amc by adding a generator-level cut on $\theta$ as $\theta > 10^{-5}$ rad.
Since this value is safely below the saturation value for the differential cross section $2 m_e / \sqrt{s}$ in the whole range of energies considered in this work, the effect of this cut on the magnitude of the cross section  is negligible.
Furthermore, the associated cross section also presents an infrared divergence from soft photon emission when $\sqrt{s} \sim m_V$, which is not present in the above formula since we assumed $\sqrt{s} \gg m_V$. This second divergence  formally cancels against the infrared divergence of the virtual 1-loop correction to the resonant production process, and represents therefore an higher order effect. 
That is, formally the events with a soft photon represent a QED radiative 
correction to the resonantly-produced dark photon.
Since we are already simulating the tree-level resonant process, we imposed $\sqrt{s} > m_V/0.95$ at the generator-level, independently of the emission angle $\theta$, to ensure that only events with sufficiently hard photons are simulated.

 We have included in our numerical evaluation the associated production rate, while we leave for future refinements  the estimation of the LDM signal arising from Compton-like dark photon production. As pointed out in the main text, we expect this process to be sizeable only in the limited region where the dark photons are massive enough to suppress bremsstrahlung, but light enough so that resonant production is not available due to the experimental energy thresholds.

\newpage

\bibliographystyle{utphys}
\bibliography{biblio.bib}

\end{document}